\documentclass[traditabstract]{aa}

\usepackage{natbib}
\bibpunct{(}{)}{;}{a}{}{,} 
\usepackage{amssymb}
\usepackage{url}
\usepackage{graphicx}
\usepackage{longtable}
\usepackage[usenames,dvipsnames]{color}
\usepackage{comment}

\newcommand{\submm}{submillimetre}

\newcommand{\smgs}{{\submm} galaxies}

\newcommand{\urltt}[1]{\texttt{#1}}
\newcommand{\micron}{\mbox{$\mu$m}}
\newcommand{\msun}{\mbox{$M_\odot$}}
\newcommand{\lsun}{\mbox{$L_\odot$}}
\newcommand{\msunyr}{\mbox{\msun\ yr$^{-1}$}}
\newcommand{\hi}{\ion{H}{i}}
\newcommand{\mhi}{\mbox{$M_{\rm H{\sc I}}$}}
\newcommand{\zhi}{\mbox{$z_{\rm HI}$}}
\newcommand{\lphi}{\mbox{$L'_{\rm HI}$}}

\newcommand{\mstar}{\mbox{$M_{*}$}}
\newcommand{\kms}{\mbox{km~s$^{-1}$}}
\newcommand{\htwo}{\mbox{H$_2$}}
\newcommand{\mhtwo}{\mbox{$M_{\rm H_2}$}}

\newcommand{\GiveRef}[1]{\citetalias{#1}: \citet{#1}}

\newlength{\propwidth}
\begin{document}

 \title{Massive stars formed in atomic hydrogen reservoirs: \\ {\hi} observations of gamma-ray burst host galaxies}
 
\titlerunning{Neutral hydrogen in GRB hosts}
\authorrunning{Micha{\l}owski et al.}

\author{Micha{\l}~J.~Micha{\l}owski\inst{\ref{inst:roe}}
\and
G.~Gentile\inst{\ref{inst:gent},\ref{inst:brus}}	
\and
J.~Hjorth\inst{\ref{inst:dark}}				
\and
M.~R.~Krumholz\inst{\ref{inst:krum}}			
\and
N.~R.~Tanvir\inst{\ref{inst:tan}}				
\and
P.~Kamphuis\inst{\ref{inst:atnf}}			
\and
D.~Burlon\inst{\ref{inst:sydney}}			
\and
M.~Baes\inst{\ref{inst:gent}}				
\and
S.~Basa\inst{\ref{inst:basa}}  				
\and
S.~Berta\inst{\ref{inst:mpe}}				
\and
J.~M.~Castro~Cer\'{o}n\inst{\ref{inst:jm}} 		
\and
D.~Crosby\inst{\ref{inst:roe}} 				
\and
V.~D'Elia\inst{\ref{inst:elia1},\ref{inst:elia2}} 	
\and
J.~Elliott\inst{\ref{inst:mpe}}				
\and
J.~Greiner\inst{\ref{inst:mpe}}				
\and
L.~K.~Hunt\inst{\ref{inst:hunt}}				
\and
S.~Klose\inst{\ref{inst:taut}}				
\and
M.~P.~Koprowski\inst{\ref{inst:roe}}			
\and
E.~Le Floc'h\inst{\ref{inst:sacley}} 			
\and
D.~Malesani\inst{\ref{inst:dark}}			
\and
T.~Murphy\inst{\ref{inst:sydney}} 			
\and
A.~Nicuesa Guelbenzu\inst{\ref{inst:taut}} 	
\and
E.~Palazzi\inst{\ref{inst:pal}}				
\and
J.~Rasmussen\inst{\ref{inst:dark},\ref{inst:dtu}}	
\and
A.~Rossi\inst{\ref{inst:pal},\ref{inst:taut}}		
\and 
S.~Savaglio\inst{\ref{inst:calab},\ref{inst:eso}}	
\and
P.~Schady\inst{\ref{inst:mpe}}				
\and
J.~Sollerman\inst{\ref{inst:soll}}				
\and
A.~de Ugarte Postigo\inst{\ref{inst:ant},\ref{inst:dark}}	
\and
D.~Watson\inst{\ref{inst:dark}}				
\and
P.~van der Werf\inst{\ref{inst:vdw}}			
\and
S.~D.~Vergani\inst{\ref{inst:meudon},\ref{inst:cov}}	
\and
D.~Xu\inst{\ref{inst:dark}} 					
	}

\institute{
SUPA\thanks{Scottish Universities Physics Alliance}, Institute for Astronomy, University of Edinburgh, Royal Observatory, Blackford Hill, Edinburgh, EH9 3HJ, UK, 
\label{inst:roe} 
e-mail: {\tt mm@roe.ac.uk}
\and
Sterrenkundig Observatorium, Universiteit Gent, Krijgslaan 281-S9, 9000, Gent, Belgium  \label{inst:gent}
\and
Department of Physics and Astrophysics, Vrije Universiteit Brussel, Pleinlaan 2, 1050 Brussels, Belgium \label{inst:brus}
\and
Dark Cosmology Centre, Niels Bohr Institute, University of Copenhagen, Juliane Maries Vej 30, DK-2100 Copenhagen \O, Denmark  \label{inst:dark}
\and
Department of Astronomy and Astrophysics, University of California, Santa Cruz, CA 95064, USA \label{inst:krum}
\and
Department of Physics and Astronomy, University of Leicester, University Road, Leicester, LE1 7RH, UK \label{inst:tan}
\and
CSIRO Astronomy \& Space Science, Australia Telescope National Facility, PO Box 76, Epping, NSW 1710, Australia\label{inst:atnf}
\and
Sydney Institute for Astronomy, School of Physics, The University of Sydney, NSW 2006, Australia \label{inst:sydney}
\and
Aix Marseille Universit\'e, CNRS, LAM (Laboratoire d'Astrophysique de Marseille) UMR 7326, 13388, Marseille, France \label{inst:basa}
\and
Max-Planck-Institut f\"{u}r Extraterrestrische Physik, Giessenbachstra{\ss}e, D-85748 Garching bei M\"{u}nchen, Germany \label{inst:mpe}
\and
ISDEFE for the Herschel Science Centre (ESA-ESAC), E-28.692 Villanueva de la Ca\~{n}ada (Madrid), Spain \label{inst:jm}
\and
ASI-Science Data Center, Via del Politecnico snc, I-00133 Rome, Italy \label{inst:elia1}
\and
INAF - Osservatorio Astronomico di Roma, Via di Frascati, 33, 00040 Monteporzio Catone, Italy \label{inst:elia2}
\and
INAF-Osservatorio Astrofisico di Arcetri, Largo E. Fermi 5, I-50125 Firenze, Italy \label{inst:hunt}
\and
Th\"uringer Landessternwarte Tautenburg, Sternwarte 5, D-07778 Tautenburg, Germany \label{inst:taut}
\and
Laboratoire AIM-Paris-Saclay, CEA/DSM/Irfu - CNRS - Universit\'e Paris Diderot, CE-Saclay, pt courrier 131, F-91191 Gif-sur-Yvette, France \label{inst:sacley}
\and
INAF-IASF Bologna, Via Gobetti 101, I-40129 Bologna, Italy \label{inst:pal}
\and
Technical University of Denmark, Department of Physics, Fysikvej, Building 309, DK-2800 Kgs.~Lyngby, Denmark \label{inst:dtu}
\and
Physics Department, University of Calabria, via P. Bucci, I-87036 Arcavacata di Rende, Italy \label{inst:calab}
\and
European Southern Observatory, Karl-Schwarzschild-Str. 2 , 85748 Garching bei M\"{u}nchen, Germany \label{inst:eso}
\and
The Oskar Klein Centre, Department of Astronomy, AlbaNova, Stockholm University, 106 91 Stockholm, Sweden \label{inst:soll} 
\and
Instituto de Astrof\' isica de Andaluc\' ia (IAA-CSIC), Glorieta de la Astronom\' ia s/n, E-18008, Granada, Spain \label{inst:ant}
\and
Leiden Observatory, Leiden University, P.O. Box 9513, NL-2300 RA Leiden, The Netherlands \label{inst:vdw}
\and
GEPI-Observatoire de Paris Meudon. 5 Place Jules Jannsen, F-92195, Meudon, France \label{inst:meudon}
\and
INAF/Osservatorio Astronomico di Brera, via Emilio Bianchi 46, 23807 Merate (LC), Italy \label{inst:cov}
}

\abstract{
Long gamma-ray bursts (GRBs), among the most energetic events in the Universe,  are explosions of massive and short-lived stars, so they  pinpoint locations of recent star formation.
However, several GRB host galaxies have recently been found to be deficient in molecular gas (\htwo), believed to be the fuel of star formation. Moreover, 
 optical spectroscopy of GRB afterglows implies that the molecular phase constitutes only a small fraction of the gas along the GRB line-of-sight.
Here we report  the first ever 21 cm line observations of GRB host galaxies, using the Australia Telescope Compact Array, implying high levels of  atomic hydrogen ({\hi}), which suggests that the connection between atomic gas and star formation is stronger than previously thought, with star formation being potentially
directly fuelled by atomic gas (or with very efficient \hi-to-{\htwo} conversion and rapid exhaustion of molecular gas), as has been theoretically shown to be possible.
This can happen in low metallicity gas near the onset of star formation, because cooling of gas (necessary for star formation) is faster than the {\hi}-to-{\htwo} conversion. 
Indeed, large atomic gas reservoirs, together with low molecular gas masses, stellar and dust masses are consistent with GRB hosts being preferentially galaxies which have very recently started a star formation episode after accreting metal-poor gas from the intergalactic medium.
This provides a natural route for forming GRBs in low-metallicity environments.
 The gas inflow scenario is also consistent with the existence of the companion {\hi} object with no optical counterpart $\sim19$\,kpc from the GRB\,060505 host, and 
 with the fact that the {\hi} centroids of the GRB\,980425 and 060505 hosts do not coincide with optical centres of these galaxies, but are located close to the GRB positions.
}


\maketitle

\section{Introduction}
\label{sec:intro}

Long (duration $>2$\,s) gamma ray-bursts (GRBs) have been shown to be related to the collapse of massive stars \citep[e.g.][]{hjorthnature,stanek}. Because of the very short main-sequence lifetime of such stars, GRBs are expected to trace galaxies with on-going star formation. Whether GRBs and their hosts can be used as an unbiased tool to trace star formation in the Universe \citep{yuksel08,kistler09,butler10,elliott12,robertson12} is still a subject of debate. Some results suggest that GRB hosts are consistent with the general population of galaxies \citep{michalowski12grb,hunt14,schady14, greiner15,kohn15}, but some biases, especially in stellar masses have been found at lower redshifts \citep{perley13,perley15,perley16,perley16b,boissier13,vergani15,schulze15}.

In order to establish the link between GRBs and star formation it is necessary to understand the properties of the gas content in the GRB hosts, which is the fuel of star formation.
The neutral hydrogen in the interstellar gas of host galaxies has been routinely detected in absorption in the spectra of GRB afterglows. In this case, the detected column density probes the portion of the neutral gas in the host galaxy which is distributed in one sight line in front of the GRB. An {\hi} column density larger than $10^{20}$ cm$^{-2}$  (a damped Lyman $\alpha$ - DLA) is measured in more than 70 GRB spectra. It was clear early on \citep{berger06,fynbo06b,savaglio06b, prochaska07b,prochaska07,watson07} that these absorbers were on average stronger than what is typically seen in quasar spectra, also probing the neutral gas in high-$z$ galaxies. It was suggested that this is because GRBs happen inside star-forming regions of the galaxy, where the  gas density is higher than in  a galaxy halo \citep{fynbo08,pontzen10}. The halo contains most of the galaxies' volume, so a sight line towards a distant quasar is likely to intercept it, rather than a dense star-forming region.
The other important difference with high-$z$ quasar-DLAs is an average higher metallicity \citep{prochaska09b,savaglio12}.

Molecular hydrogen was also detected  in absorption in a few GRB afterglow spectra.  
These data suggest low molecular hydrogen content for GRB hosts, as the measured molecular gas column density fractions ($2N(\htwo)/(2N(\htwo) + N(\hi)$) are lower than $10^{-4}$  (\citealt{vreeswijk04,fynbo06b,tumlinson07,delia10,delia14}) and are at most a few percent \citep{prochaska09,kruhler13,friis15}.

However, in order to measure the total gas content of GRB host galaxies (as opposed to column densities along a single line-of-sight  from GRB afterglow absorption spectroscopy), atomic and molecular lines need to be detected in emission.  After numerous unsuccessful searches \citep{kohno05,endo07,hatsukade07,hatsukade11b,stanway11},  the carbon monoxide (CO) line emission from three $z\sim0.1$--$0.8$ GRB host galaxies was detected and claimed to be weak compared with their star formation rates (SFRs) and dust masses \citep{hatsukade14,stanway15}. This may be surprising, because molecular gas is believed to be the fuel of star formation \citep{carilli13}, and the short duration of the main-sequence phase of GRB progenitors implies that their hosts are currently star-forming (e.g.~\citealt{christensen04,castroceron06,castroceron10,lefloch,savaglio09,levesque10c,svensson10,kruhler11,kruhler12b,hjorth12,jakobsson12,milvangjensen12,perley13,perley15,hunt14,schady14}; but see \citealt{rossi14}), and this process should be fuelled by molecular gas. 

However, a large fraction of the interstellar medium (ISM) is in the form of neutral hydrogen. This is especially true for dwarf galaxies \citep[e.g.][]{hunt14b}, and to some extent for moderate-mass galaxies ($\mstar\sim10^{10}$--$10^{11}\,\msun$; \citealt{leroy08,lagos11}), which dominate the cosmic SFR density  \citep{brinchmann04}, so we expect them to host a large fraction of GRBs.  Up to date, there has been no systematic study of the molecular and neutral gas content of GRB hosts, and of its relation to their (usually) moderate SFRs and dust masses.

The objectives of this paper are to: {\it i)} present the first detection of the neutral  hydrogen gas emission in a sample of GRB host galaxies; {\it ii)} test whether the amounts of gas are at odds with what is usually found in other star-forming galaxies; and {\it iii)} consider what the atomic gas masses of GRB hosts can tell us about the fuelling of star formation and the evolutionary stage of GRB hosts.

We use a cosmological model with $H_0=70$ km s$^{-1}$ Mpc$^{-1}$,  $\Omega_\Lambda=0.7$, and $\Omega_m=0.3$ and the \citet{chabrier03}  initial mass function (IMF), to which all SFR and stellar masses were converted (by dividing by 1.8) if given originally assuming the \citet{salpeter} IMF.

\section{Samples and Data}
\label{sec:data}
\label{sec:mass}

\defcitealias{tinney98}{1}
\defcitealias{prochaska04}{2}
\defcitealias{ofek06gcn}{3}
\defcitealias{vergani10gcn}{4}
\defcitealias{starling11}{5}
\defcitealias{levan11gcn}{6}
\defcitealias{michalowski14}{7}
\defcitealias{watson11}{8}
\defcitealias{michalowski09}{9}
\defcitealias{castroceron10}{10}
\defcitealias{thone08}{11}
\defcitealias{hatsukade07}{12}
\defcitealias{sollerman05}{13}
\defcitealias{levesque10c}{14}
\defcitealias{levesque11}{15}
\defcitealias{michalowski15rad}{16}
\begin{table*}
\scriptsize
\centering
\caption{Properties of our sample of GRB hosts.\label{tab:sample}}
\begin{tabular}{llccccccccccccc}
\hline
GRB & $z_{\rm opt}$ & Ref & SFR$_{\rm IR/Rad}$ & Ref & SFR$_{\rm UV}$ & Ref & $\log(M_*/M_\odot)$   & Ref & $\log(M_d/M_\odot)$   & Ref  & $\log(\mhtwo/M_\odot)$\tablefootmark{a} & Ref & Met & Ref\\
    &              & & ($M_\odot\mbox{ yr}^{-1}$) & & ($M_\odot\mbox{ yr}^{-1}$)   \\
\hline
980425 & 0.0085 & \citetalias{tinney98} & $0.26$ & \citetalias{michalowski14} & $$0.14 & \citetalias{michalowski09} & 8.68 & \citetalias{michalowski14} & $6.57$ & \citetalias{michalowski14} & $<8.05$ & \citetalias{hatsukade07} & 8.60 & \citetalias{sollerman05} \\
031203 & 0.1050 & \citetalias{prochaska04} & $2.8$ & \citetalias{watson11} & $$2.4 & \citetalias{castroceron10} & 9.47 & \citetalias{castroceron10} & $<8.00$\tablefootmark{d} & \citetalias{watson11} & $ \cdots $ &  $\cdots$  & 8.27 & \citetalias{levesque10c} \\
060505 & 0.0889 & \citetalias{ofek06gcn} & $0.69$ & $\ddagger$ & $$1.1 & \citetalias{castroceron10} & 9.64 & \citetalias{thone08} & $ \cdots $ &  $\cdots$  & $ \cdots $ &  $\cdots$  & 8.30 & \citetalias{thone08} \\
100316D & 0.0591 & \citetalias{vergani10gcn},\citetalias{starling11} & $1.7$ & $\ddagger$ & $$0.30\tablefootmark{b} & \citetalias{starling11} & 8.93\tablefootmark{c} & $\ddagger$ & $ \cdots $ &  $\cdots$  & $ \cdots $ &  $\cdots$  & 8.30 & \citetalias{levesque11} \\
111005A & 0.0133 & $\ddagger$,\citetalias{levan11gcn} & $0.42$ & $\ddagger$ & $$0.16\tablefootmark{b} & $\ddagger$ & 9.68 & $\ddagger$ & $6.57$ & $\ddagger$ & $ \cdots $ &  $\cdots$  & 8.70 & \citetalias{michalowski15rad} \\
\hline
\end{tabular}
\tablefoot{ \scriptsize
Unless otherwise stated, the properties are adopted from the references given in the Ref columns. The metallicities are $12+O/H$, where the solar value is 8.69 \citep{asplund04}.
\tablefoottext{a}{Converted to the CO-to-{\htwo} factor of $\alpha_{\rm CO}=5$.}
\tablefoottext{b}{We calculated the SFR from the ultraviolet flux using the conversion of \citet{kennicutt}.}
\tablefoottext{c}{We calculated the stellar mass from ultraviolet, optical and near-infrared fluxes reported in \citet{starling11}, \citet{cano11b} and \citet{olivares12} using {\sc Grasil} (see Sec.~\ref{sec:mass}).}
\tablefoottext{d}{We calculated the dust mass using the submm flux limit and assuming $T_d=30$ K and $\beta=1.5$.}
}
References: $\ddagger$: this work,  \GiveRef{tinney98},  \GiveRef{prochaska04}, 
\GiveRef{ofek06gcn},  \GiveRef{vergani10gcn},  \GiveRef{starling11}, 
\GiveRef{levan11gcn},  \GiveRef{michalowski14},  \GiveRef{watson11}, 
\GiveRef{michalowski09},  \GiveRef{castroceron10},  \GiveRef{thone08}, 
\GiveRef{hatsukade07},  \GiveRef{sollerman05},  \GiveRef{levesque10c}, 
\GiveRef{levesque11},  \GiveRef{michalowski15rad}, 
\end{table*}

\subsection{GRB hosts}
\label{sec:datahi}

\begin{table*}
\centering
\caption{ATCA radio observation details.\label{tab:radioobs}}
\begin{tabular}{lccccccc}
\hline
GRB & Date & $t_{\rm int}$\tablefootmark{a} & \multicolumn{5}{c}{Beam size ($''$)\tablefootmark{b}} \\
\cline{4-8}
         &           &          (hr)                     & 1.38  & 1.86 & 2.35 & 2.80 & {\hi}  \\
\hline
980425 &  12 Apr 2012 & 12/12 & $23\phantom{.0}\times14\phantom{.0}$ & $13.5\times8.3$ & $12.0\times7.5$ & $10.4\times7.0$ & $36\times22$		\\ 
031203 & 20--22 Jul 2013, 11,14 Apr 2014 & 38.5/21 & $\phantom{1}8.4\times\phantom{1}5.3$ & $\phantom{1}6.1\times 3.9$ & $\phantom{1}4.7\times 3.1$ & $\phantom{1}4.0\times 2.7$ &   $13\times\phantom{2}5$ \\ 
060505 &  18,19,23  Jul 2013, 11--14 Apr 2014 & 43.5/25 & $13.4\times \phantom{1}4.8$ & $\phantom{1}9.4\times 3.4$ & $\phantom{1}7.2\times 2.7$ & $\phantom{1}6.1\times 2.4$ &   $19\times\phantom{2}6$ \\ 
100316D & 22--25 Jul 2013, 12,13 Apr 2014 & 40/19 & $\phantom{1}6.5\times \phantom{1}5.1$ & $\phantom{1}4.7\times 3.7$ & $\phantom{1}3.7\times 2.9$ & $\phantom{1}3.2\times 2.5$ &   $\phantom{2}8\times\phantom{2}7$ \\ 
111005A & 18,25 Jul 2013, 11-14 Apr 2014 & 13.5/5.5 & $29\phantom{.0}\times\phantom{1}4.3$ & $17.7\times3.3$ & $13.5\times2.6$ & $11.6\times2.2$ &  $314\times16$\tablefootmark{c} \\ 
\hline
\end{tabular}
\tablefoot{
\tablefoottext{a}{The first number is the total integration time on source, and the second is the time when the {\hi} line was covered.}
\tablefoottext{b}{At the frequency in GHz specified in the header, or at the frequency of the {\hi} line.}
\tablefoottext{c}{The {\hi} data for the GRB\,111005A host was adopted from  \citet{theureau98} and \citet{springob05}.}
}
\end{table*}

Our {\hi} target sample comprises all five $z<0.12$ GRB hosts in the southern hemisphere. This ensures that they are visible by the Australia Telescope Compact Array (ATCA) and that  the {\hi} line is in the accessible frequency range.
Table~\ref{tab:sample} shows the basic properties of these hosts, whereas Table~\ref{tab:radioobs} lists the details of the ATCA observations.

We performed radio observations with ATCA using the Compact Array Broad-band Backend \citep[CABB;][]{cabb} on 12 Apr 2012, 18--25 Jul 2013 and 11--14 Apr 2014  (project no.~C2700, PI: M.~Micha{\l}owski). 
The array was in the 1.5B configuration with baselines up to 1286 m for the GRB\,980425 host and in the 6A configuration with baselines up to 5939 m for other hosts. 
The data reduction and analysis were done using the {\sc Miriad} package \citep{miriad,miriad2}.

One intermediate frequency (IF) was centred at 2.1\,GHz  with a 2\,GHz bandwidth (2048 channels 1 MHz wide each). 
We analysed the data separately in four $0.5$ GHz frequency ranges centred at 1324, 1836, 2348, and 2860 MHz. For the GRB\,980425 host we used a $50$\arcsec\  diameter aperture 
to measure the fluxes. Other hosts are not resolved, so we applied Gaussian fitting.
The continuum data for the GRB\,980425 host was presented in \citet{michalowski14}.

A second IF was centred at the {\hi} line in the ATCA CABB `zoom' mode with 32 kHz resolution. We subtracted the continuum to
obtain the continuum-free data, and made the Fourier inversion to get a
data cube with a velocity resolution of 6.6 km s$^{-1}$. 
We then made a CLEAN deconvolution down to $\sim$ 3$\sigma$, after which we restored the
channel maps with a Gaussian beam.

\begin{figure*}
\begin{center}
\begin{tabular}{cc}
\includegraphics[width=0.45\textwidth,clip]{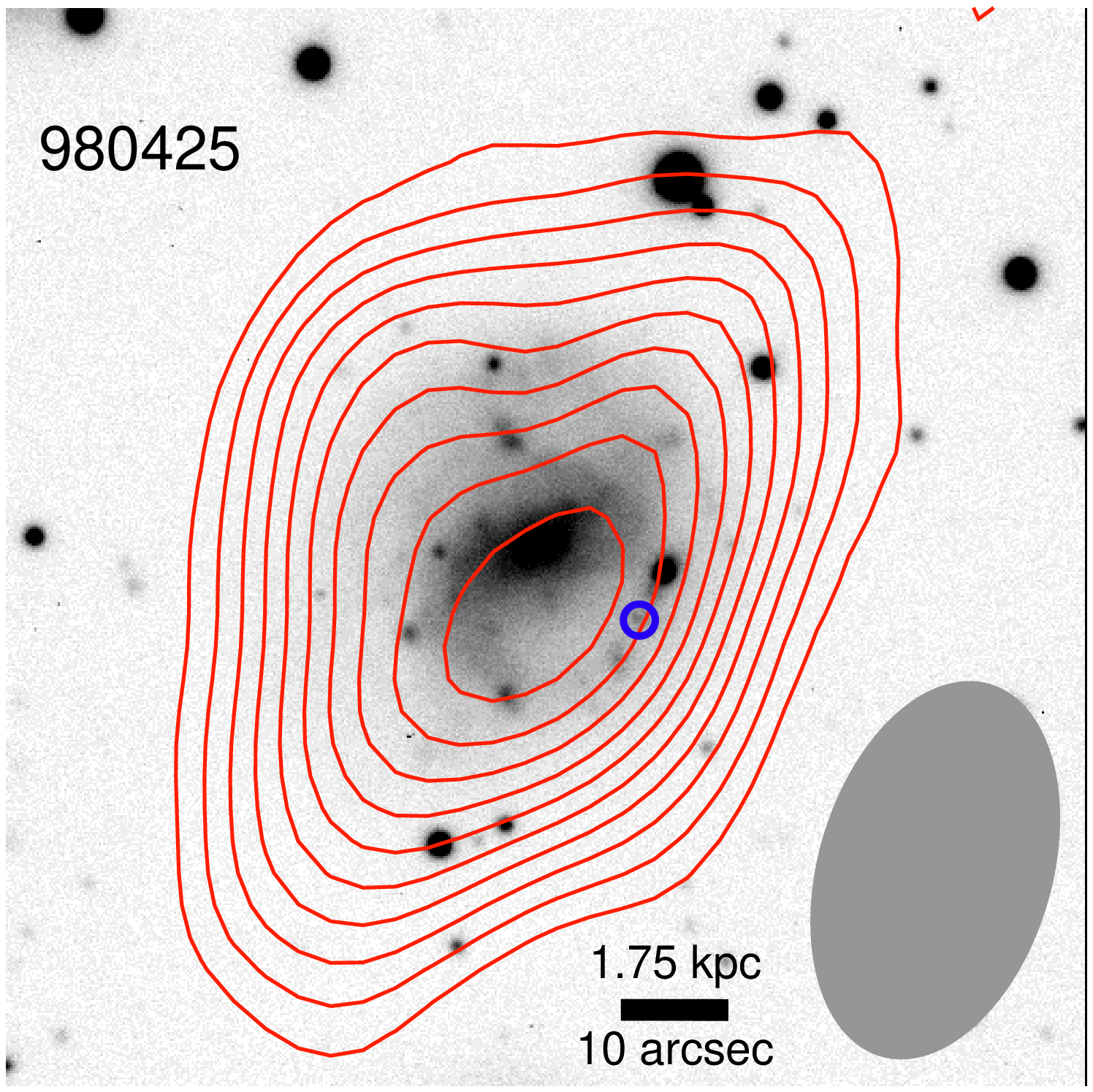} &
\includegraphics[width=0.45\textwidth,clip]{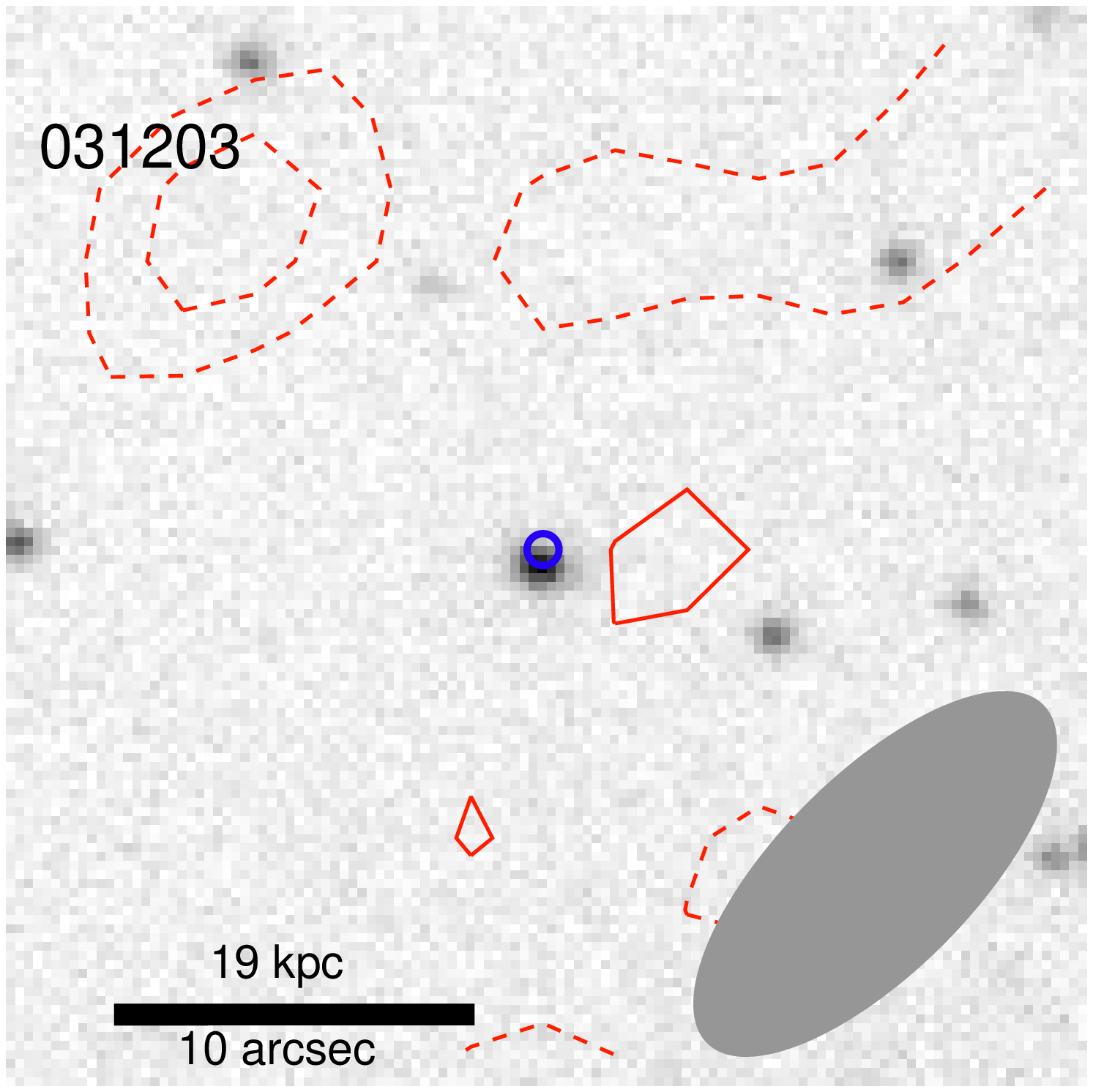} \\ 
\includegraphics[width=0.45\textwidth,clip]{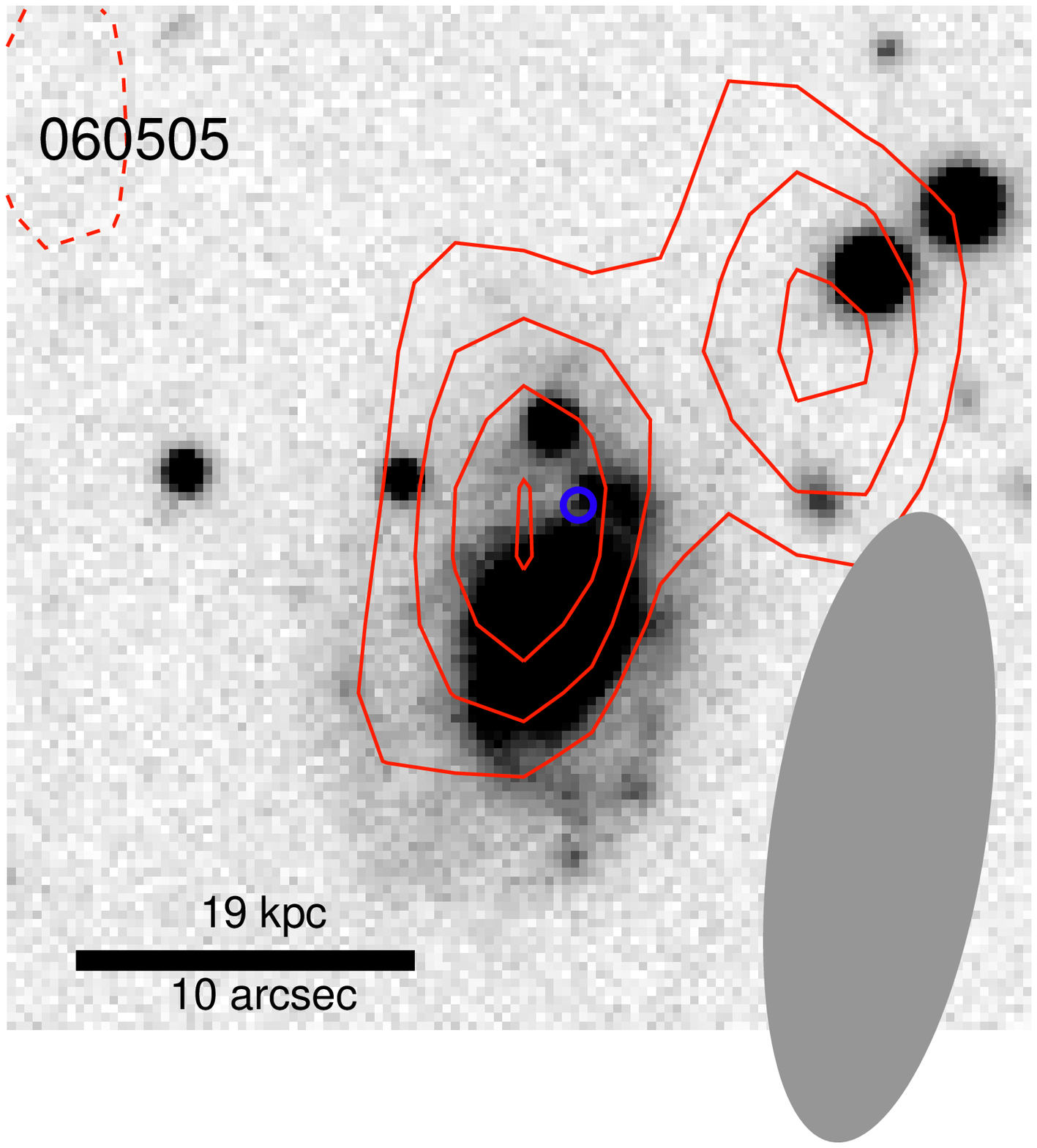} &
\includegraphics[width=0.45\textwidth,clip]{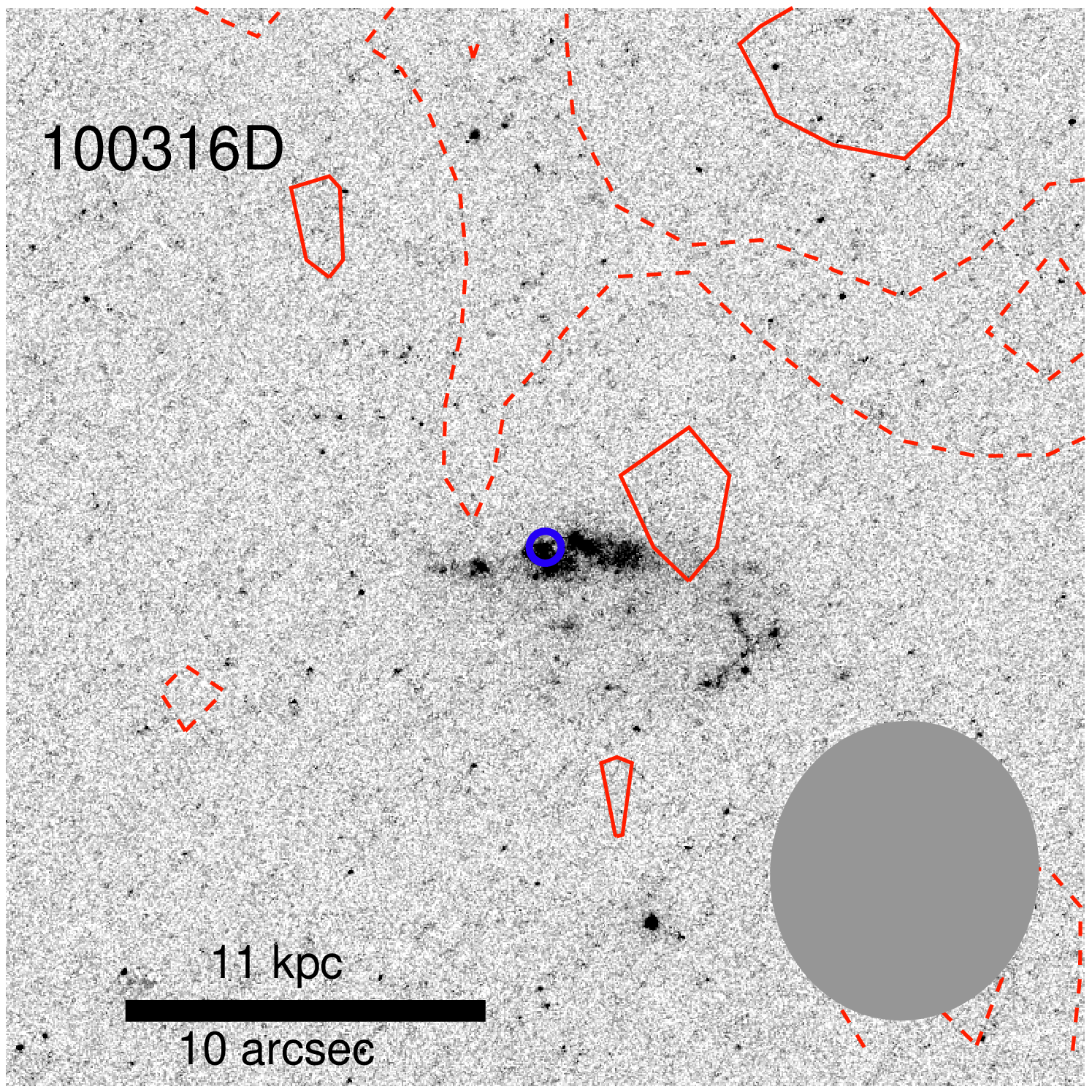} \\
\end{tabular}
\end{center}
\caption{Integrated {\hi} line map of the GRB hosts ({\it contours}; positive/negative values are solid/dashed at $-2$, $-1$, $2$, $3$, $4\sigma$, \dots) on top of the optical images \citep{sollerman05,mazzali06,thone08,starling11}. The spectrum was integrated in the frequency range shown as dotted lines in Fig.~\ref{fig:hispec}. North is up and east is to the left. The images are  $30''\times30''$ except for GRB\,980425, for which it is  $100''\times100''$. The scale is indicated on each panel. The FWHM beam sizes of the {\hi} data are shown as {\it filled ellipses}. The {\it blue  circles}  show the GRB positions. The map for the GRB\,111005A host is not shown, as we use archival data with poor spatial resolution (Sec.~\ref {sec:datahi}). 
To estimate the {\hi} properties for the GRB\,060505 host only the central source was used. The north-western source is at the same redshift (Fig.~\ref{fig:hispec}) and is analysed separately (Table~\ref{tab:mhi} and Sec.~\ref{sec:hipos}).
}
\label{fig:hiim}
\end{figure*}

\newlength{\panelwidth}
\setlength{\panelwidth}{0.5\textwidth}

\begin{figure*}
\begin{center}
\begin{tabular}{cc}
\includegraphics[width= \panelwidth]{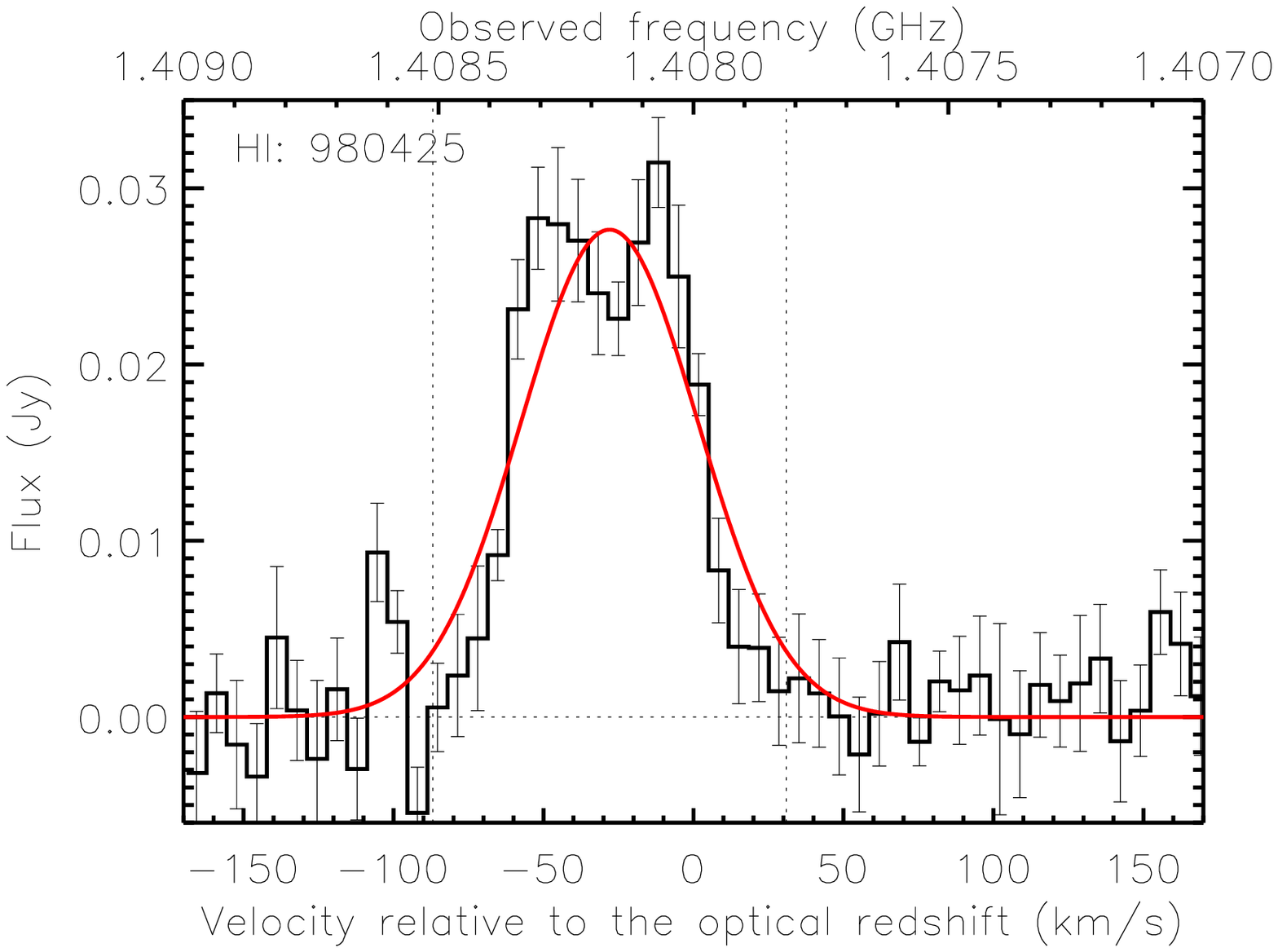} &
\hspace{-2em}%
\includegraphics[width= \panelwidth]{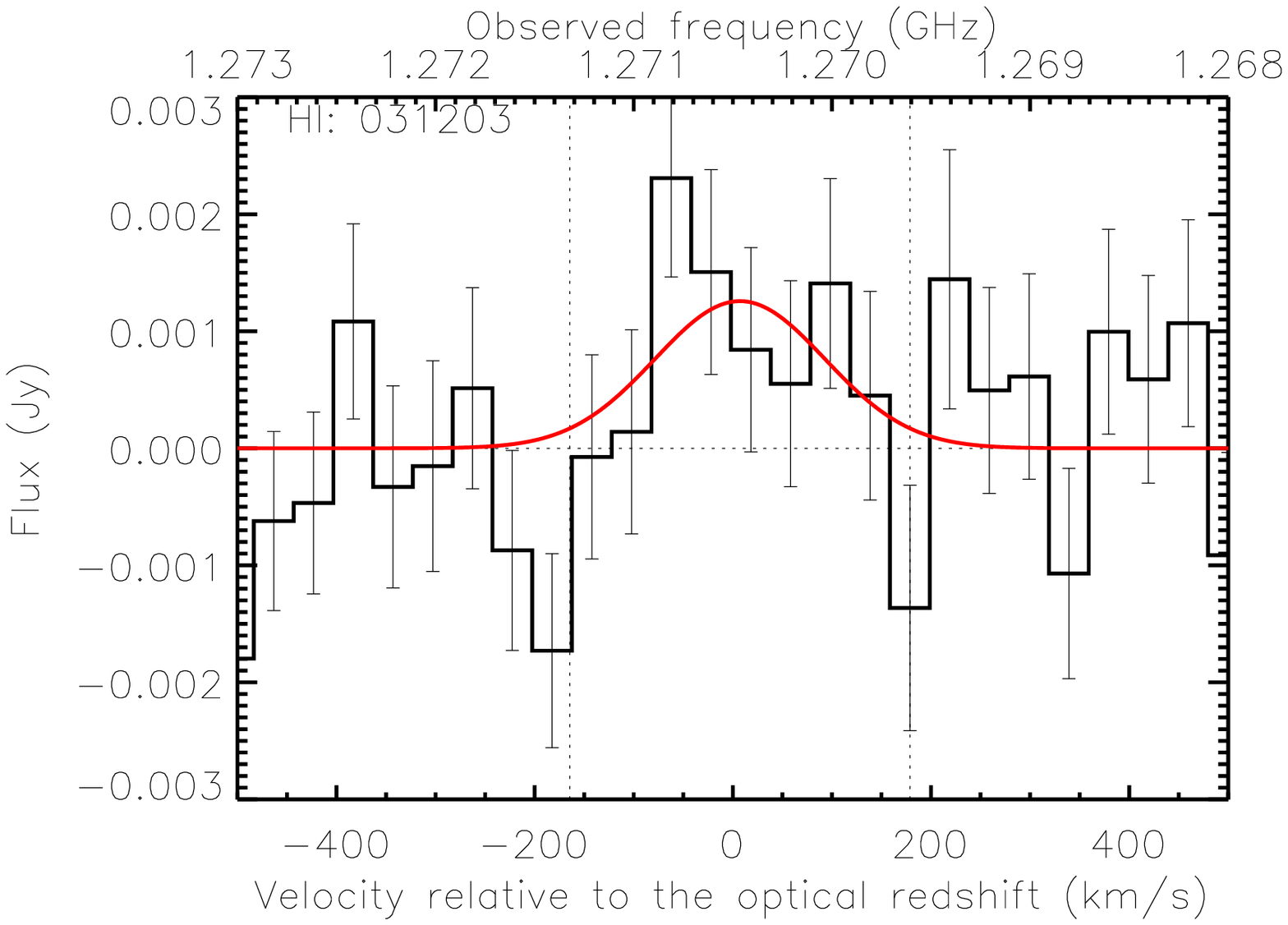} \\ 
\includegraphics[width= \panelwidth]{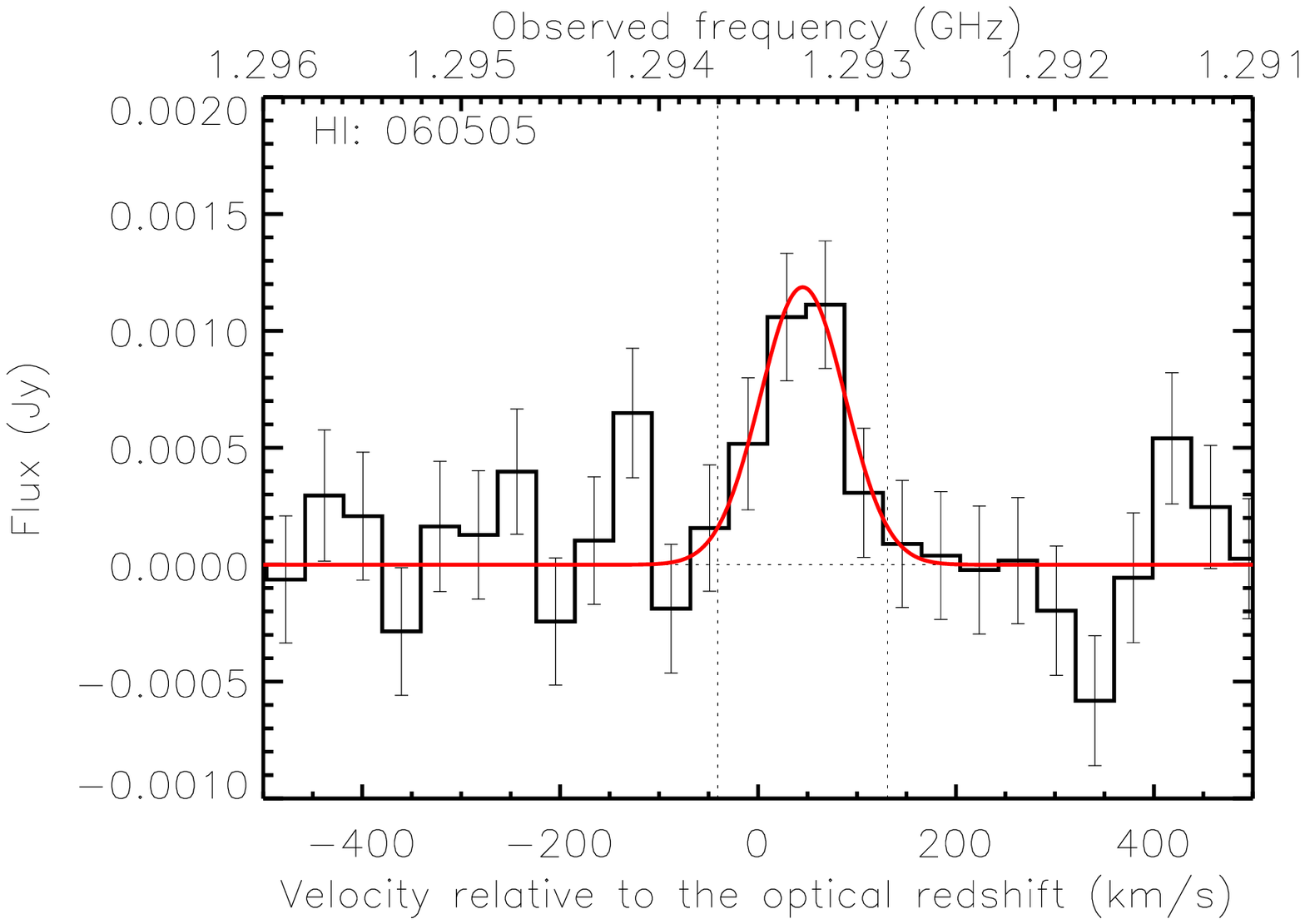} &
\hspace{-2em}\includegraphics[width= \panelwidth]{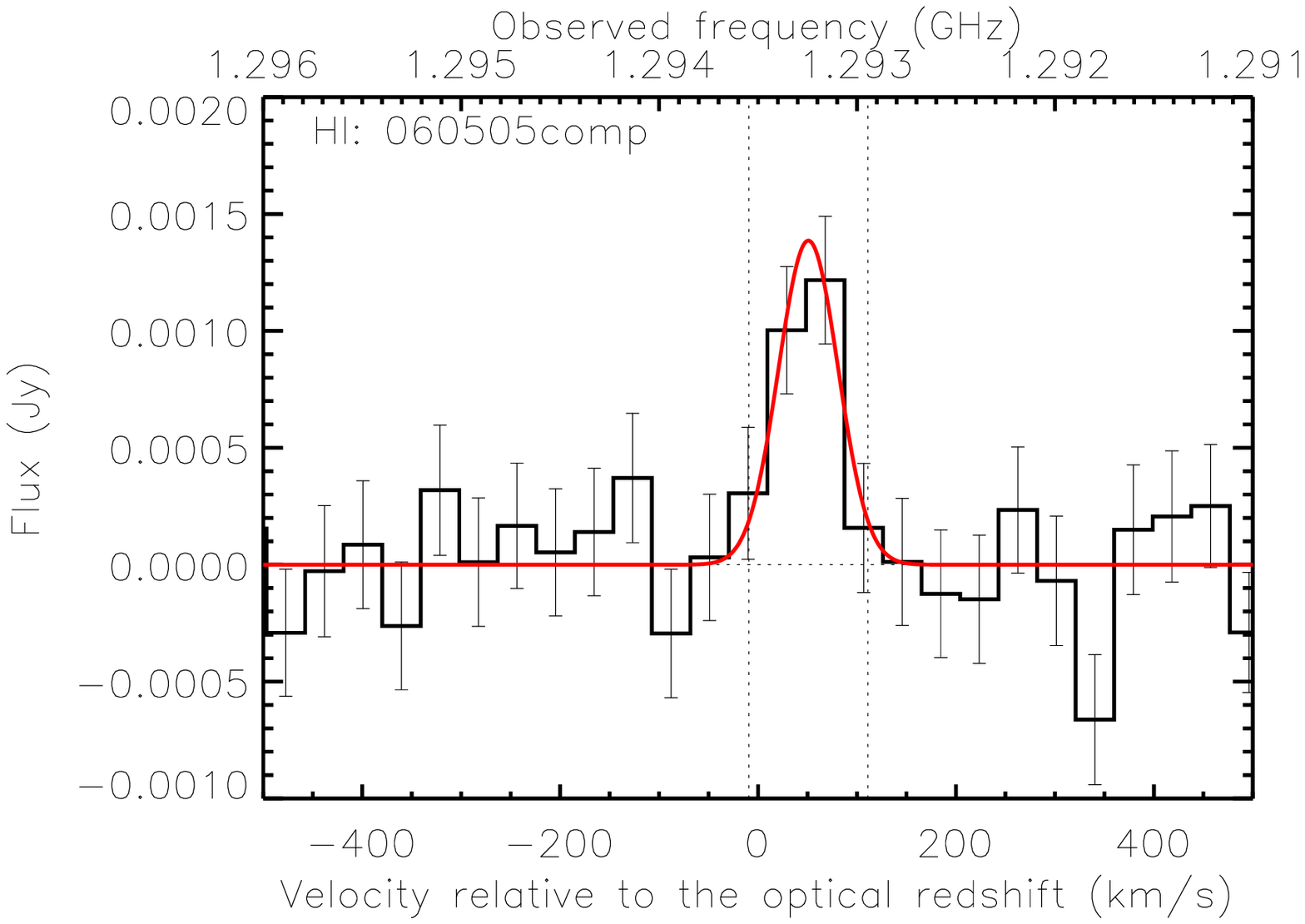} \\
\includegraphics[width= \panelwidth]{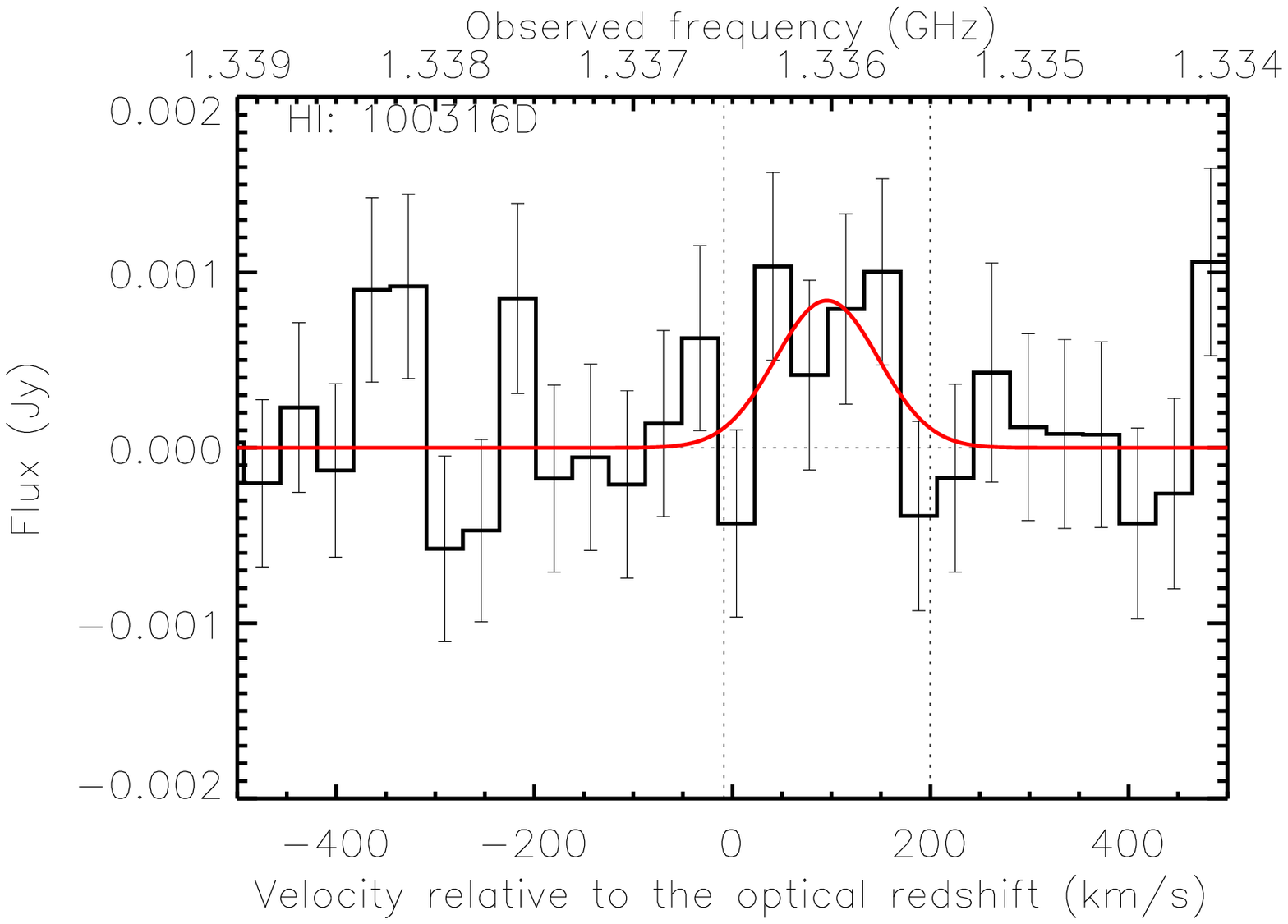} &
\includegraphics[width= \panelwidth]{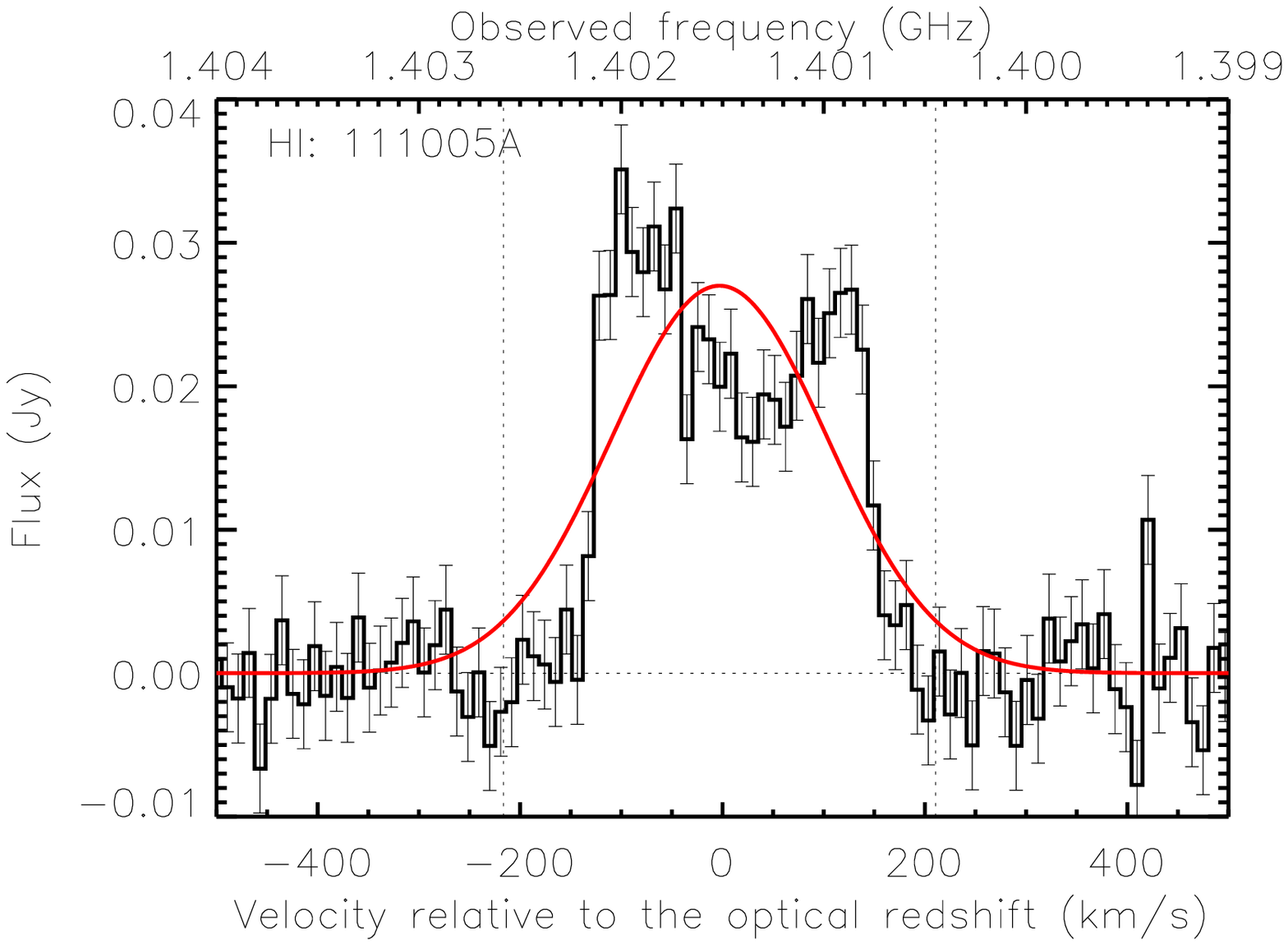}\\
\end{tabular}
\end{center}
\caption{{\hi} spectra of the GRB hosts ({\it histograms}) with the Gaussian fit overplotted ({\it red lines}). These fits are presented only for illustration, as the total fluxes were determined from the direct integration of the spectra (Sec.~\ref{sec:reshi}). {\it Vertical dotted lines} show the frequency range used to measure the total {\hi} emission and to produce the integrated {\hi} maps (Fig.~\ref{fig:hiim}). The spectrum marked `060505comp' corresponds to the companion object north-west of the GRB\,060505 position (Fig.~\ref{fig:hiim}).
}
\label{fig:hispec}
\end{figure*}

\begin{table}
\centering
\caption{The host photometry for the GRB\,100316D host using the GROND data from \citet{olivares12}. \label{tab:100316d}}
\begin{tabular}{rc}
\hline
Wavelength & mag$_{\rm AB}$  \\
($\mu$m) & \\
\hline
0.45870 & 18.09 $\pm$ 0.03\\
0.62198 & 17.85 $\pm$ 0.03\\
0.76407 & 17.75 $\pm$ 0.03 \\
0.89896 & 17.66 $\pm$ 0.04\\
1.23992 & 17.49 $\pm$ 0.05\\
1.64684 & 17.49 $\pm$ 0.06\\
2.17055 & 17.78 $\pm$ 0.09\\
\hline
\end{tabular}
\end{table}

\begin{table*}
\small
\centering
\caption{{\hi} properties of GRB hosts.\label{tab:mhi}}
\medskip
\begin{tabular}{lcccccc}
\hline
GRB & \zhi & $v_{\rm FWHM}$ & $F_{\rm peak}$ & $F_{\rm int}$   &  $\log(\lphi)$          & $\log(\mhi)$ \\
    &      & (km s$^{-1}$)  & (mJy)           & (Jy km s$^{-1}$) & ($\mbox{K km s}^{-1} \mbox{ pc}^2$) & (${\rm M}_\odot$) \\
(1)    & (2)     & (3)  & (4)           & (5) & (6) & (7) \\
\hline
980425 & $0.008607 \pm        0.000005$ & $69 \pm 4$ & $27.6 \pm 1.0$ & $2.17 \pm 0.09$ & $10.679 \pm 0.018$ & $8.849 \pm 0.018$   \\
031203 & $\cdots$ & $\cdots$ & $ < 2.2$ & $ < 0.69$ & $ < 12.404$ & $ < 10.613$   \\
060505 & $0.089508 \pm        0.000034$ & $101 \pm 32$ & $1.2 \pm 0.2$ & $0.15 \pm 0.03$ & $11.577 \pm 0.074$ & $9.780 \pm 0.074$   \\
060505comp & $0.089527 \pm        0.000025$ & $71 \pm 15$ & $1.4 \pm 0.2$ & $0.14 \pm 0.02$ & $11.531 \pm 0.072$ & $9.734 \pm 0.072$   \\
100316D & $\cdots$ & $\cdots$ & $ < 1.5$ & $ < 0.30$ & $ < 11.540$ & $ < 9.731$   \\
111005A & $0.013229 \pm        0.000014$ & $251 \pm 10$ & $27.0 \pm 0.9$ & $7.75 \pm 0.25$ & $11.606 \pm 0.014$ & $9.778 \pm 0.014$   \\
\hline
\end{tabular}
\tablefoot{
(1) GRB number. (2) Redshift determined from the Gaussian fit to the {\hi} spectrum. (3) Full width at half maximum of this Gaussian. (4) Peak of this Gaussian. (5) Integrated flux within $2\sigma$ of the Gaussian width. (6) {\hi} line luminosity using equation 3 in \citet{solomon97}. (7) Neutral hydrogen mass  using equation 2 in \citet{devereux90}. The row marked `060505comp' corresponds to the companion object Northwest of the GRB\,060505 position (Fig.~\ref{fig:hiim}).
}
\end{table*}

For the GRB\,111005A host we used  archival Nan\c cay telescope {\hi} data from \citet{theureau98} and \citet{springob05}. This galaxy was not recognised as a GRB host at the time of these observations. The galaxy appears point-like in these data, because of the beam size of $4'\times20'$ \citep{springob05}. The spectrum has a channel width of $25$ kHz  ($\sim5\,\kms$).

The potential association of GRB\,111005A with the  galaxy ESO 580-49 at $z= 0.01326$ was suggested by \citet{levan11gcn}, and  confirmed by our multi-facility campaign  \citep{xu11gcn,xu11gcn2,michalowski11gcn,michalowski15rad}. We measured its UV emission on the maps from the {\it GALEX} \citep{galex1,galex2}\footnote{{\it Galaxy Evolution Explorer}; \url{http://galex.stsci.edu/}} archive, obtaining fluxes of $524\pm	 38\,\mu$Jy and $79\pm20\,\mu$Jy at the NUV and FUV filters, respectively.

 We calculated the stellar mass of the GRB\,100316D host using the photometry from \citet{starling11}, \citet{cano11b} and the photometry from the data presented in \citet{olivares12}. The latter measurements are presented in Table~\ref{tab:100316d}. We applied the SED fitting method detailed in \citet[][see therein a discussion of the derivation of galaxy properties and typical uncertainties]{michalowski08,michalowski09,michalowski10smg,michalowski10smg4,michalowski12mass,michalowski14mass} based on 35\,000 templates in the library of \citet{iglesias07}, plus additional templates of \citet{silva98} and \citet{michalowski08}, all developed in {\sc Grasil}\footnote{\url{www.adlibitum.oat.ts.astro.it/silva}} \citep{silva98}. They are based on numerical calculations of radiative transfer within a galaxy that is assumed to be a triaxial system with diffuse dust and dense molecular clouds, in which stars are born.

For the GRB\,031203 host we calculated a dust mass upper limit from the $870\,\micron$ non-detection of \citet{watson11}, and we did not use the value of $10^{4.27}\,\msun$ calculated by \citet{symeonidis14}  based only on  detections at wavelengths shorter than $100\,\micron$, as this reflects the amount of hot dust. Cold dust (dominating the total dust mass) does not emit significantly at these wavelengths.
 Using the relation between cold and warm dust masses derived by \citet[][their Fig.~12]{izotov14}, the hot dust mass derived by \citet{symeonidis14} corresponds to the cold dust mass of $10^{6.8}\,\msun$, an order of magnitude lower than our upper limit.

Metallicities of our sample are shown in the second to last column of  Table~\ref{tab:sample}. They span the range $0.4$--$1.0$ solar.
They were derived from emission-line diagnostics, so they reflect gas-phase metallicities.

 Unless otherwise stated, when we refer to the SFRs of GRB hosts we use the infrared or radio estimates from the forth column of Table~\ref{tab:sample}.

\subsection{Other galaxy samples}
\label{sec:other}

In order to place the GRB hosts in the context of general galaxy populations we compared their properties with those of the following galaxy samples, chosen based on the availability of the gas mass estimates:
local spirals \citep{devereux90},
optical flux limited spirals and irregulars with IRAS data \citep{young89b},
local Luminous Infrared Galaxies \citep[LIRGs;][]{sanders91},
local Ultra Luminous Infrared Galaxies   \citep[ULIRGs;][]{solomon97},
the {\it Herschel} Reference Survey \citep[HRS;][]{boselli10,cortese12b,cortese14,boselli14,ciesla14},
{\hi}-selected $z<0.02$ galaxies \citep{doyle06},
{\hi}-rich galaxies \citep{wang13},
the GALEX Arecibo SDSS Survey (GASS) $0.025 < z< 0.05$ galaxies with $\log(M_*/M_\odot)>10$ \citep{catinella10},
the COLD GASS survey supplementing CO data \citep{saintonge11},
The H I Nearby Galaxy Survey \citep[THINGS;][]{walter08},
 {\hi}-dominated, low-mass galaxies and large spiral galaxies \citep{leroy08},
$0.01<z<0.03$ mass-selected galaxies with $8.5<\log(\mstar/\msun)<10$ \citep{bothwell15},
dwarf Local Irregulars That Trace Luminosity Extremes (LITTLE) THINGS \citep{hunter12},
local dwarfs  \citep{stilp13},
{\it Herschel} Virgo Cluster Survey \citep[HeViCS;][]{davies10},
 Blue Compact dwarfs \citep{grossi10},
{\hi}-selected Arecibo Legacy Fast ALFA \citep[ALFALFA;][]{giovanelli05} dwarfs \citep{huang12},
a volume-limited sample of dwarfs at distances $<4$ Mpc \citep{ott12},
metal-poor dwarfs \citep{hunt14b,leroy07},
metal-poor dwarfs from the {\it Herschel} Dwarf Galaxy Survey \citep{cormier14},
$z\sim1.5$ BzK galaxies \citep{daddi10,magdis11,magnelli12b},
and $1.2<z<4.1$ {\smgs} \citep{bothwell13,michalowski10smg}.

All dust mass estimates were converted to a common dust mass absorption coefficient $\kappa_{850\,\micron}=0.35$ cm$^2$ g$^{-1}$, all stellar masses and SFRs were converted to the \citet{chabrier03} IMF, and all molecular masses were converted to $\alpha_{\rm CO}=5\,M_\odot\, (\mbox{K km s}^{-1} \mbox{ pc}^2)^{-1}$.
 This Galactic value is appropriate for $0.4$--$1$ solar metallicity galaxies discussed here \citep{bolatto13,hunt14b}.

 SFR estimates of other galaxies are often derived from various diagnostics (UV, H$\alpha$, IR, radio), but they were shown to be broadly consistent \citep{salim07,wijesinghe11}, even in dwarf galaxies, except of very low $\mbox{SFR}<0.001\msunyr$ \citep{huang12,lee09b}, not discussed here.

\section{Results}
\label{sec:res}

\subsection{Neutral hydrogen {\hi} line}
\label{sec:reshi}

\begin{figure*}
\begin{center}
\includegraphics[width= \propwidth]{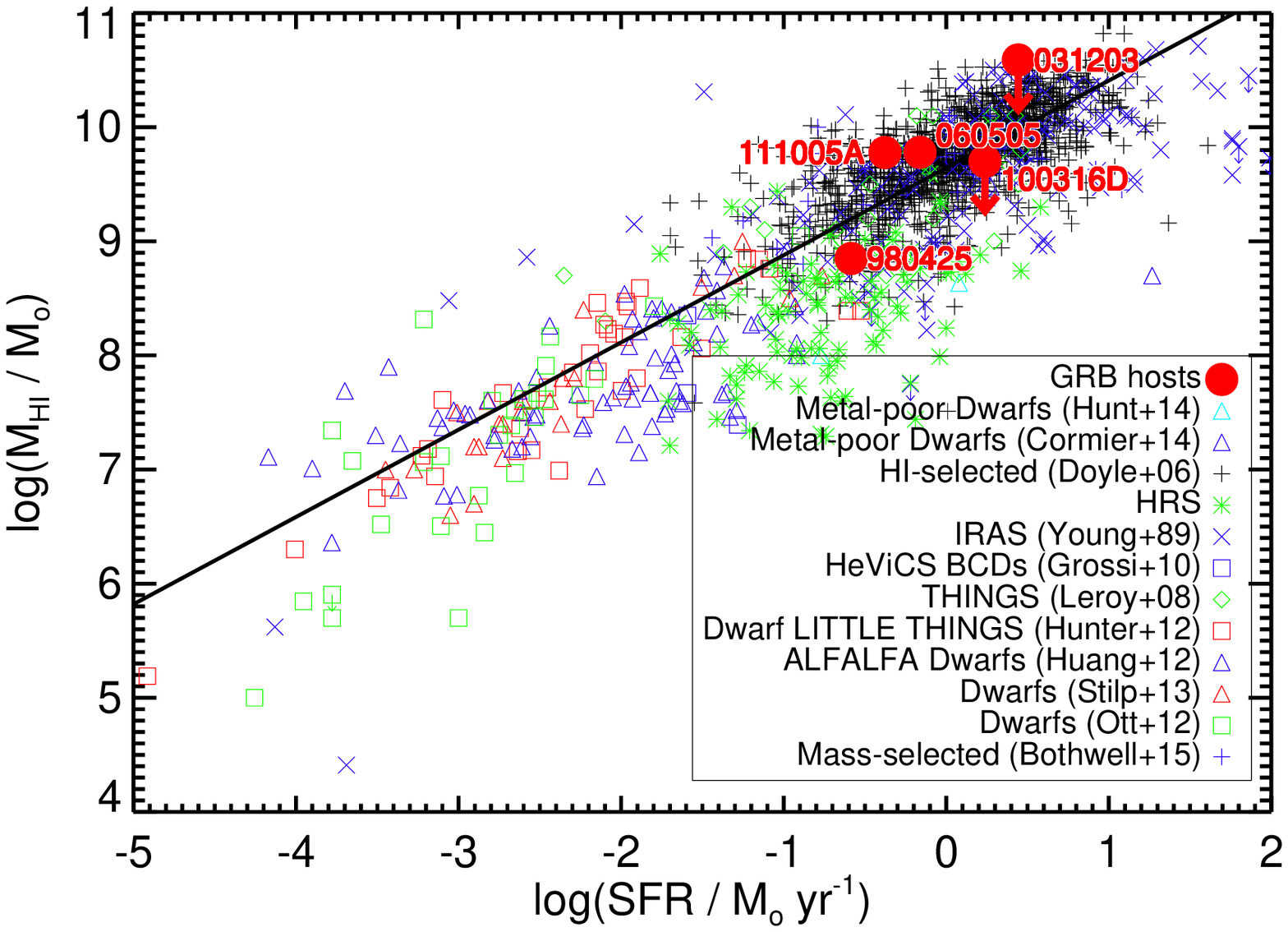}
\end{center}
\caption{{\hi} mass as a function of infrared/radio-based star formation rate (SFR) of GRB hosts ({\it red circles}) and other galaxies, as indicated in the legend and described in Sect.~\ref{sec:other}. The {\it solid black line} is a linear fit to all the data (eq.~\ref{eq:SFRMHI}). GRB hosts are consistent with the general star-forming galaxy population.
}
\label{fig:SFRMHI}
\end{figure*}

\begin{figure*}
\begin{center}
\includegraphics[width= \propwidth]{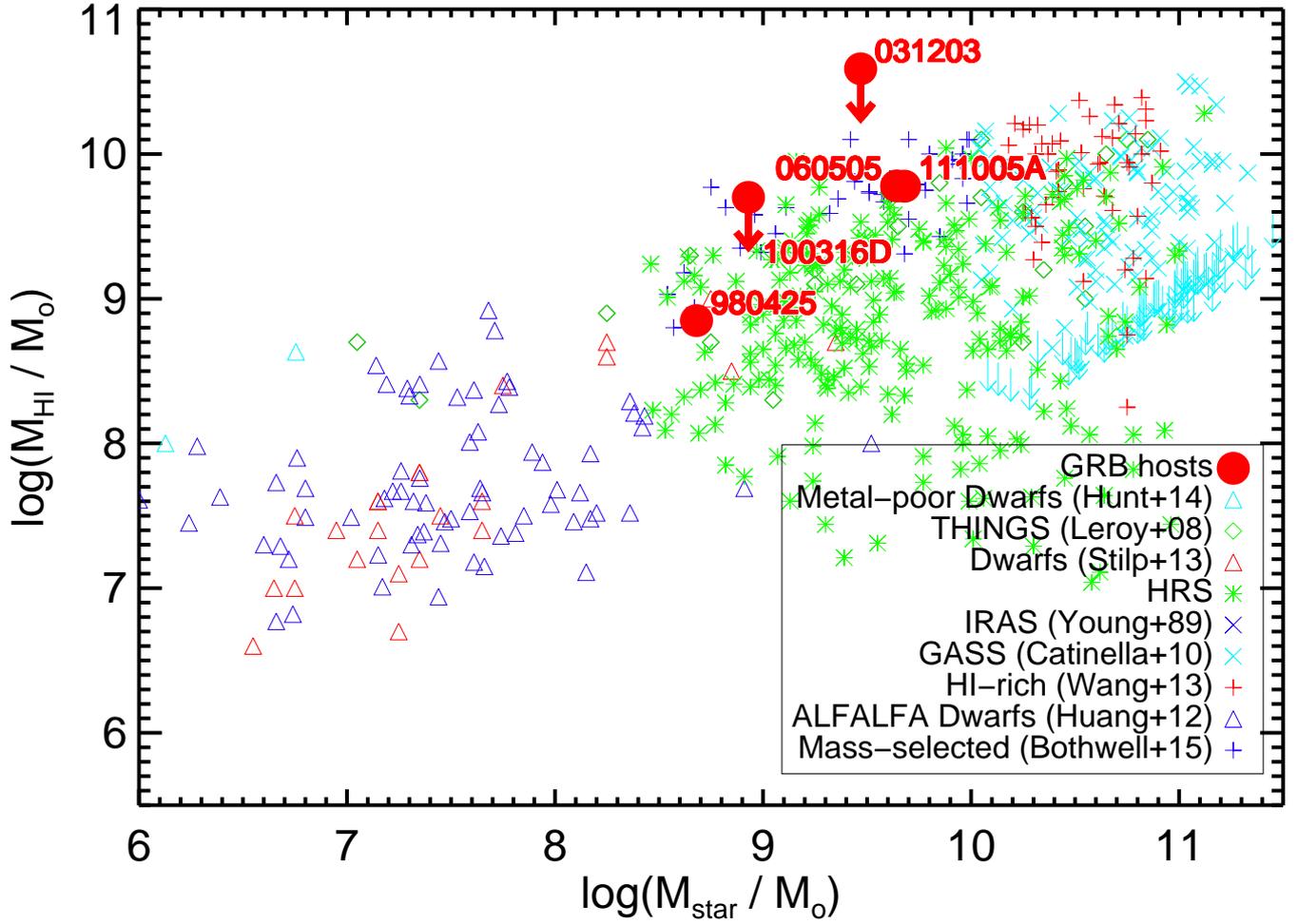}
\end{center}
\caption{{\hi} mass as a function of stellar mass of GRB hosts ({\it red circles}) and other galaxies, as indicated in the legend and described in Sect.~\ref{sec:other}. 
GRB hosts are consistent with the general star-forming galaxy population. 
}
\label{fig:MsMHI}
\end{figure*}

\begin{figure*}
\begin{center}
\includegraphics[width= \propwidth]{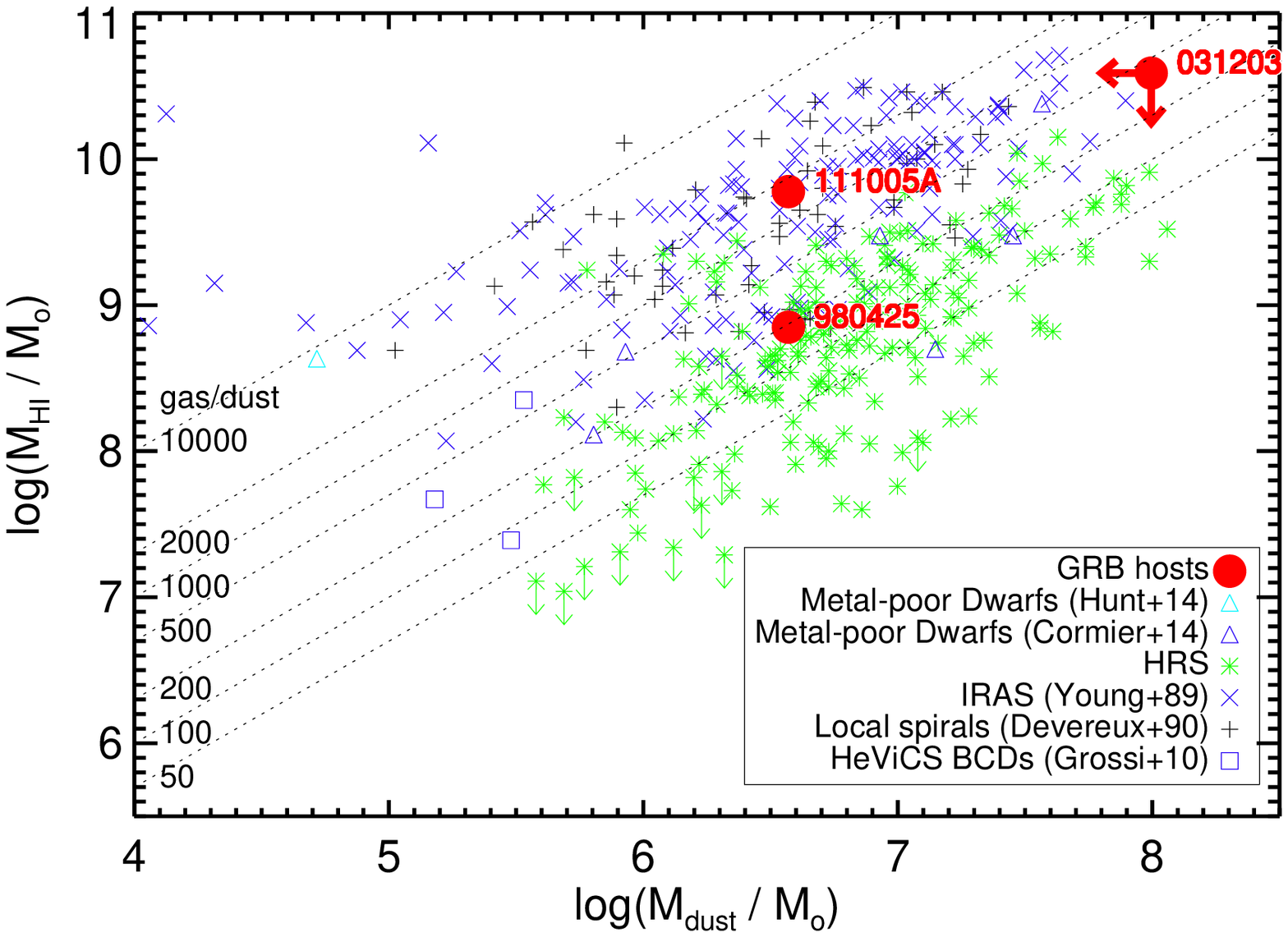}
\end{center}
\caption{{\hi} mass as a function of dust mass of GRB hosts ({\it red circles}) and other galaxies, as indicated in the legend and described in Sect.~\ref{sec:other}. 
Dotted lines denote constant gas-to-dust mass ratios with values indicated on the left.
GRB hosts are consistent with the general star-forming galaxy population, especially with the {\it Herschel} Reference Survey (HRS) galaxies with similarly measured dust masses.
}
\label{fig:MdMHI}
\end{figure*}

The fluxes at each frequency element were determined by aperture photometry with the aperture radius of $75''$ for the GRB\,980425 (to encompass the entire {\hi} emission, see Fig.~\ref{fig:hiim}) and of $8''$ for the remaining targets. For the GRB\,111005A the spectrum was directly available from \citet{theureau98} and \citet{springob05}. Gaussian functions were fitted to the spectra (Fig.~\ref{fig:hispec}) and the parameters of the fit are reported in columns 2--4 of Table~\ref{tab:mhi}.  The {\hi} emission map derived from the collapsed cube within $2\sigma$ from this fit 
 (dotted lines on Fig.~\ref{fig:hispec})
is shown on  Fig.~\ref{fig:hiim}. This range was also used to obtain integrated {\hi} emission ($F_{\rm int}$ in Jy\,\kms) directly from the spectra (not from the Gaussian fit, which in some cases is not a perfect representation of the line shape). The line luminosity ({\lphi} in K\,\kms pc$^2$) was calculated using equation 3 in \citet{solomon97} and transformed to {\mhi} using equation 2 in \citet{devereux90}. 

We detected the {\hi} emission of the hosts of GRB\,980425, 060505 and 111005A. 
This is the first time when atomic gas is detected in emission from a GRB host (see also \citealt{arabsalmani15b} on the GRB\,980425 host, based partially on the same data).
 Moreover, for GRB\,060505 we detected {\hi} emission $\sim10\arcsec$ (19\,kpc at its redshift) north-west of the centre of the host {\hi} emission ($\mbox{R.A.}=22$:07:02.9, $\mbox{Dec.}=-27$:48:46.4; source `060505comp' on Fig.~\ref{fig:hispec} and in Table~\ref{tab:mhi}). The redshift of this object is consistent with that of the GRB\,060505 host (velocity offset $\sim6\pm18\,\kms$). This object is discussed in Sec.~\ref{sec:hipos}. Its flux was not included in the estimate for the GRB\,060505 host.

 The beam sizes of the radio continuum maps and {\hi} maps are shown in Table~\ref{tab:radioobs}. Only the GRB\,980425 host is resolved, and its dynamical properties are presented in \citet{arabsalmani15b}. 
The lack of higher-resolution data does not affect our results, as we analyse the total {\hi} content, not its distribution.

The atomic hydrogen masses of GRB hosts and other galaxies as a function of their SFR, stellar mass and dust mass are shown in Figures~\ref{fig:SFRMHI}, \ref{fig:MsMHI} and \ref{fig:MdMHI}, respectively. 
We find that the GRB\,980425 host  has a molecular gas mass fraction of $\mhtwo/(\mhtwo + \mhi)<14$\%, which is within the range for other star-forming galaxies \citep[a few to a few tens percent;][]{young89b,devereux90,leroy08,saintonge11,cortese14,boselli14}. 

The {\hi} spectra of both GRB\,980425 and 111005A exhibit a double-peaked profile characteristic for a rotating disk.

For ESO  580-49  (the GRB\,111005A host) \citet{theureau98} provided total widths of the {\hi} line at 20\% and 50\% of the peak flux of  $284\pm15\,$\kms  and $272\pm10\,$\kms, respectively, slightly higher than the FWHM given in Table~\ref{tab:mhi}, due to the fact that the  Gaussian function does not represent the line profile accurately. Our estimate of the integrated flux (which does not involve assumptions on the line shape), agrees with $7.6\pm0.9\,$Jy \kms given by \citet{theureau98}.

\subsection{Radio continuum}

\begin{figure*}
\begin{center}
\includegraphics[width=\textwidth,viewport=10 400 450 830,clip]{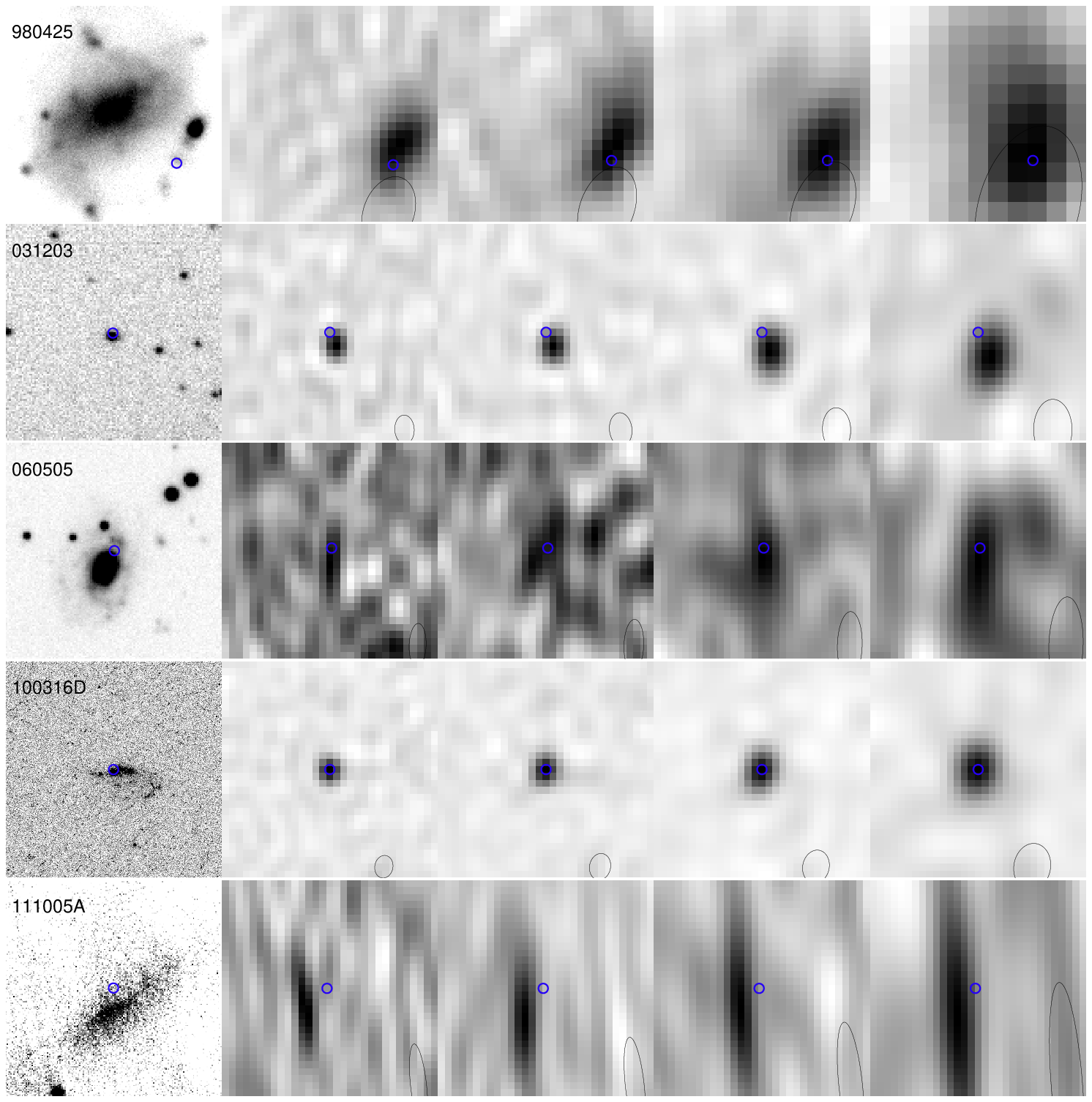}
\end{center}
\caption{Optical \citep[first column;][and this work]{sollerman05,mazzali06,thone08,starling11} and radio continuum images centred at 2800, 2340, 1860 and 1340 GHz.  North is up and east is to the left. Each panel is  $30''\times30''$ The FWHM beam size is shown on each radio panel as an {\it ellipse}. The {\it blue  circles}  show the GRB positions.
}
\label{fig:radim}
\end{figure*}

The radio images at all four frequencies are shown in Fig.~\ref{fig:radim}. All hosts are detected, but the significance for the GRB\,060505 host is only around $3\sigma$.  For each host we fitted a power law to these fluxes obtaining the radio slope $\alpha$ and the rest-frame $1.4$ GHz flux, which was transformed to $\mbox{SFR}_{\rm radio}$ using the conversion of \citet{bell03}. Finally, as in \citet{michalowski12grb}, we calculated an approximate measure of the ultraviolet (UV) dust attenuation $A_{\rm UV}=2.5\log(\mbox{SFR}_{\rm radio}/\mbox{SFR}_{\rm UV})$, which we converted to visual attenuation assuming an SMC extinction curve, which gives $A_{\rm V}=A_{\rm UV}/2.2$ \citep{gordon03}. All these estimates are reported in Table~\ref{tab:radio}. 
 Due to a similar shape of all standard extinction and attenuation curves between $0.275\,\micron$ and $V$-band, our $A_{\rm V}$ estimate would only be $\sim10$\% higher if we used the LMC or the Milky Way extinction curves \citep{gordon03}, and $\sim20$\% higher with the \citet{calzetti00} attenuation curve.

Up to date 89 long-GRB hosts have been targeted in the radio continuum \citep{bergerkulkarni,berger,vreeswijk01radio,fox03,frail03,vanderhorst05,wiersema08,michalowski09,michalowski12grb,stanway10,stanway14,watson11,perley13b,perley15}, and only 15 ($\sim17$\%) were detected: the hosts of GRB\,980425 \citep{michalowski09}, 980703 \citep{bergerkulkarni}, 000418, 010222, \citep{berger}, 021211 \citep{michalowski12grb}, 031203 \citep{stanway10,watson11,michalowski12grb}, 051022, 080207, and 090404 \citep{perley13b}; 051006, 060814, 061121, 070306,  \citep{perley15}, 080517 \citep{stanway15b}, 100621A	 \citep{stanway14}; with the addition of the hosts of short (duration $<2$\,s) GRBs 071227 \citep{nicuesa14}, and 120804A \citep{berger13}.

We provide three more detections (060505, which has been targeted before but not detected, 100316D and 111005A) bringing the fraction of detected hosts to 18/91 ($\sim20$\%).

\begin{table*}
\centering
\caption{Radio continuum properties of GRB hosts.\label{tab:radio}}
\begin{tabular}{lcccccccc}
\hline
GRB & $F_{1.38}$ & $F_{1.86}$ & $F_{2.35}$ & $F_{2.80}$ & $F_{\rm rest 1.4}$\tablefootmark{a} & $\alpha$\tablefootmark{b} & SFR$_{\rm radio}$\tablefootmark{c} & $A_V$\tablefootmark{d}  \\
    & ($\mu$Jy) & ($\mu$Jy) & ($\mu$Jy) & ($\mu$Jy) & ($\mu$Jy) & & ($M_\odot\mbox{ yr}^{-1}$) & (mag) \\
\hline
980425 & $840 \pm 160$ & $1110 \pm 210$ & $710 \pm 190$ & $400 \pm 200$ & $912 \pm 263$ & $-0.58 \pm 0.41$ & $0.11 \pm 0.03$ & $\sim0$  \\ 
031203 & $383 \pm \phantom{1}20$ & $\phantom{1}324 \pm \phantom{1}11$ & $319 \pm \phantom{1}12$ & $324 \pm \phantom{1}13$ & $363 \pm \phantom{1}26$ & $-0.19 \pm 0.09$ & $2.83 \pm 0.20$ & $0.08$  \\ 
060505 & $\phantom{1}76 \pm \phantom{1}35$ & $\phantom{1}\phantom{1}67 \pm \phantom{1}18$ & $\phantom{1}45 \pm \phantom{1}18$ & $\phantom{1}47 \pm \phantom{1}15$ & $\phantom{1}85 \pm \phantom{1}50$ & $-0.81 \pm 0.72$ & $0.69 \pm 0.40$ & $\sim0$  \\ 
100316D & $657 \pm \phantom{1}21$ & $\phantom{1}531 \pm \phantom{1}13$ & $441 \pm \phantom{1}13$ & $403 \pm \phantom{1}13$ & $681 \pm \phantom{1}33$ & $-0.71 \pm 0.06$ & $1.73 \pm 0.08$ & $0.86$  \\ 
111005A & $245 \pm \phantom{1}30$ & $\phantom{1}192 \pm \phantom{1}18$ & $160 \pm \phantom{1}16$ & $124 \pm \phantom{1}16$ & $252 \pm \phantom{1}46$ & $-0.92 \pm 0.22$ & $0.08 \pm 0.02$ & $\sim0$  \\ 
\hline
\end{tabular}
\tablefoot{
Flux densities are given at the observed frequencies indicated in GHz in the column header. 
\tablefoottext{a}{Rest-frame $1.4$ GHz flux density from the power-law fit to the data.}
\tablefoottext{b}{Radio power-law slope.}
\tablefoottext{c}{Using the conversion of \citet{bell03}.}
\tablefoottext{d}{Visual dust attenuation calculated from the ultraviolet attenuation $A_{\rm UV}=2.5\log(\mbox{SFR}_{\rm radio}/\mbox{SFR}_{\rm UV})$ assuming an SMC extinction curve, which gives $A_{\rm V}=A_{\rm UV}/2.2$ \citep{gordon03}. $A_V\sim0$ mag is reported if a formal value of $\mbox{SFR}_{\rm UV}$ exceeds the $\mbox{SFR}_{\rm radio}$.}
}
\end{table*}

\section{Discussion}
\label{sec:discussion}

\subsection{Large {\hi} reservoirs in GRB hosts: early stages of star formation}
\label{sec:highhi}

It is still an unanswered question whether GRB hosts are consistent with the general population of star-forming galaxies. 
As shown in Figs.~\ref{fig:SFRMHI}, \ref{fig:MsMHI} and \ref{fig:MdMHI}, GRB hosts at $z<0.12$ contain large atomic gas reservoirs, within the ranges expected for galaxies with similar SFRs, stellar masses and dust masses. 

We performed a  two-dimensional Kolmogorov-Smirnov test using the publicly available IDL procedures\footnote{\url{http://www.astro.washington.edu/users/yoachim/code.php}} of \citet{2dksidl}.  We compared the GRB host sample to the HI-selected galaxies \citep{doyle06} on the SFR-{\mhi} plane, and the HRS \citep{boselli10} galaxies on the \mstar-{\mhi} plane, because GRB hosts span similar SFR and {\mstar} ranges to those of these galaxies, respectively. For GRB\,031203 and 100316D  with no {\hi} detections, we assumed 20 different values of their {\mhi} between $\mstar/100$ (a conservative lower limit) and the derived {\mhi} upper limits (Table~\ref{tab:mhi}). We 
obtained $p$-values
of $11$--$79$\% for the SFR-{\mhi} plane, and $0.2$--$24$\% for the  \mstar-{\mhi} plane. This implies that GRB hosts have consistent  (or somewhat higher) atomic gas masses when being compared with other samples. If the {\hi} masses of non-detected hosts are close to the derived upper limits, then the probability for the \mstar-{\mhi} plane is $0.2$\%, corresponding to a $\sim4\sigma$ discrepancy between the GRB host and the HRS samples. In such a case, GRB hosts are all located in the high-{\mhi} end of the distribution or have low {\mstar}, which is also evident on Fig.~\ref{fig:MsMHI}. If these galaxies are well below the derived upper limits, then the GRB host population is consistent with the comparison sample.

The mass at which the contribution of local galaxies to the {\hi} mass function is the highest is $\log(\mhi/\msun)\sim9.6$, as  derived from the {\mhi} function of \citet{zwaan05b}.
This is similar to what we found for GRB hosts. 
This is consistent with numerical simulations, suggesting that galaxies with low SFRs 
(similar to low-$z$ GRB hosts) dominate the {\hi} mass density at all redshifts, including the local universe \citep[Fig.~6 of][]{lagos14}. In particular, galaxies with $\mbox{SFR=}0.1$--$5\,\msunyr$ and $0.1$--$1\,\msunyr$ contribute $\sim45$\% and $\sim30$\% to the {\hi} mass density at $z<1$.  (\citealt{lagos12, lagos14}, private communication) All our {\hi}-detected GRB hosts are in this SFR range, so they can be considered typical galaxies at low-$z$ with regards to their {\hi} content.

These considerations indicate that, at least at $z\lesssim0.1$, GRB hosts can be regarded as normal star-forming galaxies. This is consistent with the conclusion presented in \citet{michalowski12grb}, \citet{hunt14} and \citet{kohn15} based on radio and far-infrared continuum data.

To complete this picture we note that  \citet{perley13,perley15,perley16b} and \citet{vergani15} found that GRB hosts at $z<1.5$ have lower stellar masses than what would be expected from the assumption that GRBs trace the cosmic star formation activity in an unbiased way. Additionally, \citet{michalowski14,michalowski15rad} found dust masses of two low-$z$ GRB hosts to be close to the lower envelope of other galaxies.

Normal SFRs and atomic gas masses, together with low stellar and molecular gas masses  are consistent with GRB hosts being preferentially galaxies which have very recently started a star formation episode. In this scenario, GRB hosts have not had enough time yet to use their atomic gas reservoir (which is then high) and to produce stars and dust (which catalyses the molecular gas formation). 

All our targets are in the nearby Universe, so are likely different from the high-$z$ counterparts. Only with the advent of the Square Kilometre Array (SKA)  the {\hi} emission of GRB hosts (and other galaxies) at higher redshifts can be detected.

\subsection{{\htwo} vs {\hi}: what is the fuel of star formation?}

Large atomic gas content, as measured from our data, together with the molecular gas deficiency can in principle be explained in four different ways, which we discuss below, providing evidence that  the first two contradict observations and that the last one (star formation directly fuelled by  a recent accretion of atomic gas) is the most plausible (because of the metallicity considerations and the success to explain low stellar masses of GRB hosts).

First, it is possible that very strong UV radiation from the star-forming region in which a GRB progenitor is born dissipated the available molecular hydrogen \citep{hatsukade14,stanway15}. However, this is unlikely to affect the entire galaxy, but just the vicinity of the GRB. Moreover, such process would also destroy dust, whereas the molecule-poor regions close to the GRB sites are dust-rich \citep{hatsukade14}, as is the region close to the GRB\,980425 \citep{michalowski14}. 
Finally, very intense UV radiation dissipating the remaining molecular gas would imply that the surrounding of a GRB is at the end of the intense star-formation episode, which should result in an enhanced metallicity in that region, contrary to observations \citep{christensen08,modjaz08,thone08,thone14, levesque10c,levesque11,han10}.

Second, low molecular gas masses in GRB hosts were derived from the CO lines, which may not be a good tracer of molecular gas at low metallicities if dust shielding is weak \citep{bolatto13},  and if a significant amount of {\htwo} gas is CO-dark, as is the case for metal-poor dwarfs. However, 
this is not the case for GRB hosts for which low molecular gas content was claimed, because they have $0.5$--$1.0$ solar metallicity \citep{castrotirado07,graham09,levesque10b,stanway15b}. Similarly, in our sample the metallicity is not very low either, around $0.4$--$1.0$ solar \citep[][see Table~\ref{tab:sample}]{sollerman05,levesque10c,thone08,levesque11}.
Moreover, a low molecular gas content in GRB hosts has also   been found independently of CO observations, using the optical afterglow spectroscopy,  tracing molecular gas column density along the GRB line-of-sight \citep{fynbo06b,tumlinson07,kruhler13,delia14}. This is not due to molecule destruction by the GRB itself, as this can only be effective up to a few parsecs from the GRB \citep{draine02}.

Third, atomic gas  may be converted into the molecular phase and be immediately used for star formation (which explains the lack of molecular gas). However, the timescale of the {\hi}-to-{\htwo} conversion is usually longer than the cooling timescale (and the star formation timescale; \citealt{krumholz12}), so  this could happen only at approximately solar metallically, at which \hi-to-{\htwo} conversion is quicker than the  star formation timescale. Metallicities of GRB hosts and GRB sites are usually much lower (see the discussion below), so this explanation is possible only if some properties of GRB hosts make the  \hi-to-{\htwo} conversion unusually effective, which is not accounted for in the models of \citet{krumholz12}.

Finally, the fourth and our preferred option is that star formation is directly fuelled by atomic gas from the intergalactic medium. This was theoretically shown to be possible \citep{glover12,krumholz12},
and must have been the case for
the very first stars in the Universe (in the absence of dust and molecular gas).
This
can also take place
at any redshift, even in a galaxy not particularly metal-poor on average, as long as it happens at early stages of a star formation episode, when the first stars are born in a collapsing cloud formed out of newly accreted metal-poor gas. 

 Physically, this scenario rests on two points. First, observations indicate that star formation begins within roughly one free-fall time of the appearance of a cold, dense phase of the ISM (for example see the review by \citealt{Dobbs14}). Only the duration of star formation is a matter of a debate, but this is not relevant here, as we concentrate on the beginning of a star formation episode. Indeed, some observations and theoretical models indicate that star formation finishes quickly after it starts, around 1--2 free-fall times \citep{elmegreen00,hartmann01,jeffries11,reggiani11,dobbs13}, whereas other suggest that the duration of star formation can be significantly longer (several up to 10 free-fall times), though it still starts around one free-fall time \citep{tan06,krumholz06,kawamura09,fukui99}. 

Second, the timescale for the appearance of a cold phase is determined by the gas cooling time, while the appearance of {\htwo} is determined by the chemical equilibration time. At low metallicities the cooling time is longer, but still remains much shorter than the free-fall time, whereas the chemical equilibration time can be much longer \citep[][fig.~1]{krumholz12}, leading to the formation of a cold phase and the onset of star formation before gas is able to convert from {\hi} to {\htwo}.
Specifically, theoretical models \citep{krumholz12} predict that at the metallicities of $\sim 0.3$ solar 
typically found in 
GRB sites \citep[including those analysed here;][]{christensen08,modjaz08, levesque10c,levesque11,han10,thone14}
gas can cool to low temperatures in less than a free-fall time, but that full conversion to molecular gas requires close to ten free-fall times, so the first stages of star formation are fuelled by atomic gas. Moreover, the observed GRB site metallicities could even overestimate the metallicity when the GRB progenitor formed, since the metallicity will have been increased during the progenitorÕs lifetime by on-going mixing with the more metal-rich gas existing in the hosts. If so, the discrepancy between molecule-formation and free-fall timescales would be even larger. Thus our preferred scenario is that GRB progenitors potentially form in the first burst of star formation that takes place in newly accreted low-metallicity inter galactic medium (IGM) gas. The progenitor then would end its life as a GRB before there is time for substantial H$_2$ to form, explaining the paucity of H$_2$ absorption features seen in GRB afterglows, and the paucity of CO emission from the host galaxy.

Indeed recent star-formation and large {\hi} content for other galaxies (as also observed for GRB hosts) are believed to be a result of recent inflow of metal-poor gas from the intergalactic medium \citep{dave13}. 
This 
creates metal-poor regions in a galaxy (as observed in other galaxies by \citealt{cresci10,sanchezalmeida13,sanchezalmeida14}), which mix with the surrounding metal-rich gas over a relatively long timescale
($\lesssim100$ Myr; \citealt{yang12,grand15,petit15}). Indeed, GRBs were shown to explode in the most metal-poor regions of their hosts \citep{christensen08,thone08, thone14,levesque11}. 
The GRB site metallicities are $\sim0.3$ solar, which would imply that less than half of the accreted gas converts to the molecular phase after one free-fall time \citep{krumholz12}. This fraction can be lower if the accreted gas had in fact lower metallicity than the GRB site measurements, which have been increased by on-going mixing with the more metal-rich gas existing in the hosts. We note that the bulk of the star formation in these regions (at later stages) will have to be fuelled by molecular gas, when the conversion from the atomic to molecular phase is complete.

The recent accretion of atomic gas also explains high specific SFRs of GRB hosts \citep[e.g.][]{castroceron10}, because the enhanced SFR has not lasted very long, so the stellar mass of a galaxy is lower than what would be expected from its high SFR. Moreover, this mechanism directly predicts that the regions around GRBs (at which the gas is accreted) have enhanced SFRs and dust masses compared with other parts of the hosts, consistent with observations \citep{lefloch,lefloch12,christensen08,thone08,thone14,levesque11,hatsukade14,michalowski14}.

Our data therefore  is consistent with a scenario whereby GRBs are preferentially produced when low-metallicity gas accretes onto a galaxy and undergoes rapid cooling and star formation before it either forms {\htwo} or mixes with the higher-metallicity gas in the remainder of the galaxy.
 This mechanism can be tested with positional and velocity information from high-resolution {\hi} observations, which may reveal concentration of {\hi} close to metal-poor regions. 

If confirmed, this mechanism will provide
 a natural explanation of  the low-metallicity and low-{\mhtwo} preference in the framework of the GRB collapsar model, which requires that GRB progenitors have low metallicity in order to reduce the loss of mass  and angular momentum (required for launching the jet; \citealt{yoon05,yoon06,woosley06}). In contrast, at later stages of star formation molecular gas is the dominant phase in the interstellar medium, but the metals are well mixed, and gas has been further enriched, so massive stars do not end their lives as GRBs, and such metal- and molecular-rich galaxies do not become GRB hosts. 

However, this picture is complicated by the fact that some GRBs have been found in galaxies with solar or supersolar metallicities  \citep{prochaska09,levesque10b,kruhler12,savaglio12,elliott13,schulze14,stanway15b, hashimoto15,schady15}.
This can be explained in two ways. First,  these metallicity measurements were obtained for the entire galaxies, so do not rule out local metallicity decrements (as observed for other GRB hosts), predicted by the mechanism of recent metal-poor gas accretion.

Second, this high-metallicity problem is alleviated by some models, which predict the GRB preference for low-metallicity, but do not exclude metal-rich examples, for example  the model in which a GRB progenitor is a rapidly rotating Helium star created by a merger of post-main sequence stars  \citep{vandenheuvel13}. In both collapsar  and binary scenarios the low-metallicity preference (even if it is not strict) is consistent with our interpretation that GRB hosts have recently started a star formation episode.

We note that the IMF in metal-poor environments has been suggested to be top-heavy (\citealt{bate05,zhang07,marks12}; but see \citealt{myers11}). Hence, the sites of GRB explosions are promising places to look for the top-heavy IMF. This IMF would imply higher massive star (and hence GRB) production per unit star formation rate in such environments.

Summarising, our results  provide the first observational  indication of star formation fuelled by atomic gas. The link between the star formation and the atomic gas for all galaxies is visible in Fig.~\ref{fig:SFRMHI} where the {\hi} mass is  strongly correlated with SFR. The linear fit to all datapoints gives
\begin{eqnarray}
\log(\mhi/\msun)&=&(0.76\pm0.01)\times\log(\mbox{SFR}/\msunyr)\nonumber \\
&&+ (9.64\pm0.01)
\label{eq:SFRMHI}
\end{eqnarray}
The SpearmanÕs rank correlation coefficient for this relation is 0.77, indicating a significant ($\sim30\sigma$) correlation. The scatter around this correlation is $\sim0.38$ dex, which is likely a consequence of galaxies with similar SFRs being at different evolutionary stages. The correlation with similar slope was also noted using the surface densities of SFR and {\mhi} \citep{kennicutt98,bigiel10, roychowdhury14}.

 The slope of $\sim0.76$ (or $\sim1.3$ with inverted axes) is shallower than the linear (slope of unity) Schmidt-Kennicutt relation between molecular gas and SFR densities \citep{bigiel08}. This indicates that overall molecular gas is better correlated with star formation, so is its major fuel. However, in outer parts of spiral galaxies, and in {\hi}-dominated galaxies, the relation between SFR and {\hi} column density is in fact linear \citep{bigiel08,bigiel10} and the SFR-gas correlation improves when atomic gas is taken into account, not just  the molecular gas \citep[][their fig.~2]{fumagalli08}. This was interpreted as {\hi} fuelling of low-levels of star formation in the outer parts of disks. A similar mechanism may also operate  in some star-forming regions of GRB hosts, because these galaxies have low molecular gas content, and are likely observed at the beginning of a star formation episode, just after accreting atomic gas from the IGM, unlike other galaxies.

\subsection{Location of the {\hi} emission}
\label{sec:hipos}

Despite the low resolution of our {\hi} data we can infer useful information about the location of the {\hi} emission, which supports our gas inflow scenario. First, the companion {\hi} object (not a nearby galaxy, as we show below)  is only $\sim19$\,kpc from the GRB\,060505 host (Fig.~\ref{fig:hiim} and \ref{fig:hispec}), so they must be physically related (the optical radius of the GRB\,060505 host is $\sim11$\,kpc; \citealt{thone08}). With the current resolution we cannot definitely determine the nature of this relation, but
 its existence is consistent with {\hi} gas inflowing from the north-western direction,  either in a form of a star-free atomic gas cloud, or with extremely low stellar surface density  (we discuss below that the astrometry of the radio and optical images are consistent). 
 The radial velocities of the GRB\,060505 host and this cloud are identical (offset of $\sim6\pm18\,\kms$, Table~\ref{tab:mhi}), which is consistent with the scenario that the cloud is flowing directly toward the host, or that the orbit is perpendicular to the line-of-sight.
The existence of this objects may indicate that the host of GRB\,060505 is at the earliest stage of the inflow compared with other hosts, so the inflowing material is still visible outside the host, unlike for other members of our sample.
 The inflow scenario for this cloud needs to be tested by high-resolution {\hi} observation, which can reveal its morphology and velocity structure.

There are two other explanations of this feature, but they are unlikely. First it could be a signature of a major merger\footnote{This could not be a minor merger because these objects have similar {\hi} masses.} with another galaxy, but the {\hi} companion object does not have any optical counterpart  (despite high $\log(\mhi/\msun)\sim9.7$; Table~\ref{tab:mhi}), so cannot be understood as a companion galaxy, but rather a large gas cloud. Namely, it is not detected with the HST B-band image down to a (point-source) limit of $27.1$\,mag AB \citep{ofek07}. This corresponds to an absolute magnitude of $M_B>-10.93$\, mag, and a luminosity $L_B<2.9\times10^6\,\lsun$. Hence it has $\log(\mhi/L_B)>3.3$, much higher than those of dwarf galaxies \citep[from $-1.5$ to $0.75$; Fig.~1 of][]{hunter12}. This limit assumes that this object is point-like in the HST resolution, but it would need to be larger  than the HST resolution $\sim20$ times  ($\sim2\arcsec$, or $\sim3\,$kpc) to make its $\mhi/L_B$ ratio consistent with those of dwarf galaxies. But then it would be an extremely low-surface brightness object. 
 Moreover, no emission lines were found at this position in the IFU data of \citet{thone14}.
 
Second, the {\hi} companion may be ejected from the GRB\,060505 host via tidal interaction or ram pressure stripping. However, this explanation is unlikely given the lack of nearby companions and that it is not located in the high-density cluster environment (Sec.~\ref{sec:env}).

Another information supporting our gas inflow interpretation is that for both the GRB\,980425 and 060505 hosts the centre of the {\hi} emission is shifted away from the optical centre of the galaxy towards the GRB position (Fig.~\ref{fig:hiim}).  If a galaxy has a steady {\hi} disk, its center should coincide with the optical center of this galaxy. On the other hand, if a galaxy has received a significant gas inflow from one direction, then in low-resolution data the {\hi} centroid will be offset from the optical center towards this direction. Hence, the offsets for GRB\,980425 and 060505 host are consistent with an inflow of atomic gas on  the regions around the GRB positions. This will be verified with future higher-resolution observations.

To test the ATCA astrometry we imaged the phase calibrator for the GRB\,060505 observations (2149-287) obtaining the position 21:52:03.735, $-$28:28:28.256, which is only $\sim0.04\arcsec$ away from the catalogue position 21:52:03.7352, $-$28:28:28.218\footnote{\url{http://www.narrabri.atnf.csiro.au/calibrators/calibrator_database_viewcal?source=2149-287}}. We also searched for common objects in the radio and optical images.
There is no other significant object in the integrated {\hi} map (Fig.~\ref{fig:hiim}), which would be inside the optical map, but in the radio continuum image $\sim1\arcmin$ south-west of the GRB position there is an $\sim0.4\,$mJy source at the position 22:07:00.093, $-$27:49:14.89 with an optical counterpart at the position 22:07:00.133, $-$27:49:15.32. Hence, the radio map is shifted only $0.7\arcsec$ to the West with respect to the optical map.

The astrometry of the optical image \citep{thone08} was tied to the Two Micron All Sky Survey catalogue \citep{2mass,2massmain}, and is accurate to a fraction of arcsec ($ \sim0.1\arcsec$). Moreover, it is consistent with more recent astrometry from \citet{hjorth12}.

Hence we conclude that the astrometry of the optical and radio images are consistent,  and therefore the {\hi}-optical offsets for the GRB 980425 and 060505 hosts are real.

 Finally, we note that the concentration of atomic gas in the south-eastern (blue-shifted; \citealt{michalowski15rad}) part of  
the GRB\,111005A host is also hinted by its asymmetric {\hi} line profile (Fig.~\ref{fig:hispec}). The blue-shifted line part is more pronounced, so the south-eastern part contains more atomic gas. 
The line profile for the GRB\,980425 is more
symmetric, but this is expected, because its metal poor region (close
to the GRB site) has the velocity close to the systemic one
\citep{christensen08}.

\subsection{Large scale environments}
\label{sec:env}

If  GRB hosts were located in dense large-scale regions (e.~g.~close to the cores of massive groups or clusters), then our interpretation of recent metal-poor gas  inflow would be difficult to advocate, because of scarcity of metal-poor gas in such environments. However, the GRB\,980425 host has been shown to be isolated \citep{foley06}. While the GRB\,060505 host lies in the foreground of a filamentary structure a few Mpc away from a galaxy cluster \citep{thone08}, at such distances gas stripping or evaporation does not take place, and the SFR is actually enhanced \citep{porter08}.

In order to investigate the environments of other hosts in our sample we searched the NASA/IPAC Extragalactic Database (NED) for companion galaxies within 1 Mpc in projection 
and within $\pm$1500 {\kms} in redshift.
The GRB\,031203 and 100316D hosts have no nearby galaxies within these search criteria.
The GRB\,111005A host is not in any NED group catalog either, and there are only 6 nearby galaxies nearby, all $\gtrsim500\,$kpc away.

Hence,  the host galaxies are all relatively isolated in terms of large-scale environment, which makes it possible that they experienced inflows of metal-poor intergalactic gas.

\subsection{Little dust-obscured star formation}

Only for the GRB\,100316D host the radio SFR exceeds significantly the UV-derived value, implying $A_V\sim0.86$\,mag (Tables~\ref{tab:sample} and \ref{tab:radio}). For the remaining hosts the comparable SFR$_{\rm radio}$ and SFR$_{\rm UV}$ indicate that there is very little dust-obscured star formation in these hosts.  This is consistent with our interpretation that GRB hosts are at the beginning of the star-formation episode, when the dust content has not had time to accumulate to make star-forming regions optically thick.

\subsection{Radio spectral slope}
\label{sec:alpha}

The radio spectral slopes (Table~\ref{tab:radio}) of all but one hosts are consistent with  
those of star-forming galaxies both local and at high redshifts  \citep[$\sim-0.75$;][]{condon,dunne09,ibar10}. 

As suggested in \citet{michalowski12grb}, the spectral slope of the GRB\,031203 host is significantly flatter ($\sim-0.19\pm0.09$), inconsistent with the normal value for star-forming galaxies. This implies a significant free-free emission \citep[or synchrotron self-absorption;][]{condon} and, hence, a younger stellar population \citep{bressan02,cannon04,hirashitahunt06,clemens08}, because only massive stars emit Lyman continuum photons, which create \ion{H}{ii}\, regions, responsible for free-free emission. The flat slope for the GRB\,031203 host also rules out any significant  AGN contribution to the radio flux.
 This is consistent with its location in the star-forming part of the Baldwin-Phillips-Terlevich \citep[BPT;][]{bpt} diagram  \citep{watson11}.

\section{Conclusions}
\label{sec:conclusion}

We report the results of the first ever {\hi} survey of GRB host galaxies, detecting three out of five targets. Large inferred atomic gas masses, together with low molecular gas, stellar and dust masses  are consistent with GRB hosts being at the beginning of a star formation episode, after accreting metal-poor gas from the intergalactic medium. This star formation may  potentially be directly fuelled by atomic gas  (or with very efficient \hi-to-{\htwo} conversion and rapid exhaustion of molecular gas), which can happen  in low metallicity gas near the onset of star formation, because cooling of gas (necessary for star formation) is faster than the {\hi}-to-{\htwo} conversion. This provides a natural route for forming GRBs in low-metallicity environments.  The gas inflow scenario is consistent with the existence of a companion {\hi} object with no optical counterpart $\sim19$\,kpc from the GRB\,060505 host, and with the offset towards the GRB positions of the {\hi} centroids for the GRB\,980425 and 060505 hosts away from the optical centres of these galaxies.

\begin{acknowledgements}

We thank Joanna Baradziej, Stefano Covino, Pawe{\l} Micha{\l}owski, Tadeusz Micha{\l}owski, and our referee  for help with improving this paper; Robert Braun, Sarah Maddison,  Anita Titmarsh, Catarina Ubach, and Ivy Wong   for help with the ATCA observations; 
 Christina Th\"one for analysing for us her IFU data for the GRB\,060505 host;
Marianne Doyle-Pegg for providing her data; 
Claudia Lagos for providing the predictions from her model.

M.J.M.~acknowledges the support of the UK Science and Technology Facilities Council, British Council Researcher Links Travel Grant, and the hospitality at the Instituto Nacional de Astrof\'{i}sica, \'{O}ptica y Electr\'{o}nica.
The Dark Cosmology Centre is funded by the Danish National Research Foundation.
L.K.H.~is supported by the INAF PRIN 2012 grant.
S.K. and A.N.G. acknowledge support by grant DFG Kl 766/16-1.
T.M. and D.B. acknowledge the support of the Australian Research Council through grant DP110102034.
A.d.U.P. acknowledges support from the European Commission (FP7-PEOPLE-2012-CIG 322307) and from the Spanish project AYA2012-39362-C02-02.

The Australia Telescope Compact Array is part of the Australia Telescope National Facility which is funded by the Commonwealth of Australia for operation as a National Facility managed by CSIRO. 
{\em Galaxy Evolution Explorer} ({\em GALEX}) is a NASA Small Explorer, launched in 2003 April. We gratefully acknowledge NASA's support for construction, operation, and science analysis for the {\em GALEX} mission, developed in cooperation with the Centre National d'Etudes Spatiales of France and the Korean Ministry of Science and Technology. 
This research has made use of data from HRS project. HRS is a Herschel Key Programme utilising Guaranteed Time from the SPIRE instrument team, ESAC scientists and a mission scientist. The HRS data was accessed through the Herschel Database in Marseille (HeDaM - {\tt http://hedam.lam.fr}) operated by CeSAM and hosted by the Laboratoire d'Astrophysique de Marseille. 
This research has made use of 
the GHostS database (\urltt{http://www.grbhosts.org}), which is partly funded by Spitzer/NASA grant RSA Agreement No. 1287913; 
the NASA/IPAC Extragalactic Database (NED) which is operated by the Jet Propulsion Laboratory, California Institute of Technology, under contract with the National Aeronautics and Space Administration;
SAOImage DS9, developed by Smithsonian Astrophysical Observatory \citep{ds9};
the NASA's Astrophysics Data System Bibliographic Services;
and the  Edward Wright Cosmology Calculator \url{www.astro.ucla.edu/~wright/CosmoCalc.html} \citep{wrightcalc}.

\end{acknowledgements}




\end{document}